\documentclass[10pt,conference]{IEEEtran}
\usepackage[T1]{fontenc}
\usepackage{cite}
\usepackage{amsmath,amssymb,amsfonts}
\usepackage{algorithmic}
\usepackage{graphicx}
\usepackage{textcomp}
\usepackage{xcolor}
\usepackage[hyphens]{url}
\usepackage{fancyhdr}
\usepackage{hyperref}

\usepackage{adjustbox} 
\usepackage{makecell} 
\usepackage{multirow}
\usepackage{subcaption}
\DeclareCaptionFont{my9pt}{\fontsize{9pt}{10.8pt}\selectfont}
\usepackage{makecell}
\usepackage{booktabs}

\newcommand{\hpcayear}{2026}
\newcommand\blfootnote[1]{%
  \begingroup
  \renewcommand\thefootnote{}
  \footnotetext{#1}
  \addtocounter{footnote}{-1}
  \endgroup
}

\usepackage{xspace}

\def\arch{PIMphony\xspace}
\def\archp{PIMphony}

\newcommand{\hpcasubmissionnumber}{342}
\title{\arch: Overcoming Bandwidth and Capacity Inefficiency in PIM-based Long-Context LLM Inference System}

\def\hpcacameraready{} 

\newcommand\hpcaauthors{%
Hyucksung Kwon\textsuperscript{*1},
Kyungmo Koo\textsuperscript{*1},
Janghyeon Kim\textsuperscript{1},
Woongkyu Lee\textsuperscript{1},
Minjae Lee\textsuperscript{1},
Gyeonggeun Jung\textsuperscript{2},\\
Hyungdeok Lee\textsuperscript{3},
Yousub Jung\textsuperscript{3},
Jaehan Park\textsuperscript{3},
Yosub Song\textsuperscript{3},
Byeongsu Yang\textsuperscript{3},
Haerang Choi\textsuperscript{3},\\
Guhyun Kim\textsuperscript{3},
Jongsoon Won\textsuperscript{3},
Woojae Shin\textsuperscript{3},
Changhyun Kim\textsuperscript{3},
Gyeongcheol Shin\textsuperscript{3},
Yongkee Kwon\textsuperscript{3},\\
Ilkon Kim\textsuperscript{3},
Euicheol Lim\textsuperscript{3},
John Kim\textsuperscript{2},
Jungwook Choi\textsuperscript{$\ddagger$1}%
}

\newcommand\hpcaaffiliation{%
\textsuperscript{1}Hanyang University, Seoul, Republic of Korea,\\
\textsuperscript{2}KAIST, Daejeon, Republic of Korea,\\
\textsuperscript{3}Solution Advanced Technology, SK hynix, Republic of Korea%
}

\newcommand\hpcaemail{%
\footnotesize
\texttt{\{momarom, kookyungmo, kkt20, lwghanyang, Imj4666, choij\}@hanyang.ac.kr}\\
\texttt{gyeonggeun@kaist.ac.kr, jjk12@kaist.edu}\\
\texttt{\{hyungdeok.lee, ryan.song, jaehan3.park, yosub.song, byeongsu.yang, haerang.choi, guhyun.kim, jongsoon.won,}\\
\texttt{woojae.shin, changhyun4.kim, gyeongcheol.shin, yongkee.kwon, ilkon.kim, euicheol.lim\}@sk.com}\\
}

\author{
  \ifdefined\hpcacameraready
    \IEEEauthorblockN{\hpcaauthors{}}
      \IEEEauthorblockA{
        \hpcaaffiliation{} \\
        \hpcaemail{}
      }
  \else
    \IEEEauthorblockN{\normalsize{HPCA \hpcayear{} Submission
      \textbf{\#\hpcasubmissionnumber{}}} \\
      \IEEEauthorblockA{
        Confidential Draft \\
        Do NOT Distribute!!
      }
    }
  \fi 
}

\fancypagestyle{camerareadyfirstpage}{%
  \fancyhead{}
  
  \fancyhead[C]{
    \ifdefined\aeopen
    \parbox[][12mm][t]{13.5cm}{\hpcayear{} IEEE International Symposium on High-Performance Computer Architecture (HPCA)}    
    \else
      \ifdefined\aereviewed
      \parbox[][12mm][t]{13.5cm}{\hpcayear{} IEEE International Symposium on High-Performance Computer Architecture (HPCA)}
      \else
      \ifdefined\aereproduced
      \parbox[][12mm][t]{13.5cm}{\hpcayear{} IEEE International Symposium on High-Performance Computer Architecture (HPCA)}
      \else
      \parbox[][0mm][t]{13.5cm}{\hpcayear{} IEEE International Symposium on High-Performance Computer Architecture (HPCA)}
    \fi 
    \fi 
    \fi 
    \ifdefined\aeopen 
      \includegraphics[width=12mm,height=12mm]{ae-badges/open-research-objects.pdf}
    \fi 
    \ifdefined\aereviewed
      \includegraphics[width=12mm,height=12mm]{ae-badges/research-objects-reviewed.pdf}
    \fi 
    \ifdefined\aereproduced
      \includegraphics[width=12mm,height=12mm]{ae-badges/results-reproduced.pdf}
    \fi
  }
  \fancyfoot[C]{}
}
\begin{document}

\maketitle

\ifdefined\hpcacameraready 
  \thispagestyle{camerareadyfirstpage}
  \pagestyle{empty}
\else
  \thispagestyle{plain}
  \pagestyle{plain}
\fi

\newcommand{\hpcaheight}{0mm}
\ifdefined\eaopen
\renewcommand{\hpcaheight}{12mm}
\fi


\begin{abstract}

The expansion of long-context Large Language Models (LLMs) creates significant memory system challenges. While Processing-in-Memory (PIM) is a promising accelerator, we identify that it suffers from critical inefficiencies when scaled to long contexts: severe channel underutilization, performance-limiting I/O bottlenecks, and massive memory waste from static KV cache management. In this work, we propose \arch, a PIM \emph{orchestrator} that systematically resolves these issues with three co-designed techniques. First, \textit{Token-Centric PIM Partitioning (TCP)} ensures high channel utilization regardless of batch size. Second, \textit{Dynamic PIM Command Scheduling (DCS)} mitigates the I/O bottleneck by overlapping data movement and computation. Finally, a \textit{Dynamic PIM Access (DPA)} controller enables dynamic memory management to eliminate static memory waste. Implemented via an MLIR-based compiler and evaluated on a cycle-accurate simulator, \arch significantly improves throughput for long-context LLM inference (up to 72B parameters and 1M context length). Our evaluations show performance boosts of up to 11.3$\times$ on PIM-only systems and 8.4$\times$ on xPU+PIM systems, enabling more efficient deployment of LLMs in real-world long-context applications.

\end{abstract}
\blfootnote{\textsuperscript{*}Equal contribution. \textsuperscript{$\ddagger$}Corresponding author.}

\section{Introduction}

The rapid advancement of Large Language Models (LLMs) has transformed fields ranging from natural language processing to intelligent agents by enabling the generation of contextually relevant responses across diverse applications~\cite{openai2024gpt4technicalreport, comanici2025gemini25pushingfrontier, grattafiori2024llama3herdmodels, anthropic2024sonnet, jiang2023mistral7b}. In particular, long-context LLMs, capable of maintaining coherence across hundreds of thousands of tokens, have significantly enhanced contextual relevance in various tasks. For instance, long document summarization~\cite{zhong-etal-2021-qmsum} generates cohesive summaries from dispersed information across different sections of extensive text, while repository-level code analysis~\cite{liu2023repobench} extends programming assistants’ capabilities to analyze entire codebases comprising thousands of lines. Furthermore, chain-of-thought (CoT) reasoning has recently improved answer quality by leveraging multi-step contextual reasoning~\cite{yang2025qwen3technicalreport, mpt7b, anthropic2024sonnet, openai2024o1}.

As these LLMs continue to expand — some models  exceeding 1T parameters and context windows of 1M tokens~\cite{qwen3max,team2024gemini,meta2025llama4} — the memory bandwidth and capacity are known to be the bottlenecks of overall system performance~\cite{recasens2025mind,cent}. 
Conventional accelerators such as multi-GPUs suffer from poor compute utilization since large matrix-vector (GEMV) operations with low compute intensity dominate these autoregressive decoding phases~\cite{hong2023dfx}. 
Recent work such as FlashInfer~\cite{ye2025flashinfer} and others~\cite{TensorRT-LLM,vLLM,SGLang} propose different frameworks to accelerate LLM on GPU-based systems through token-level parallel processing or dynamic batch-size optimization; however, their performance benefits are still bounded by GPU's limited memory bandwidth and capacity.

Recently, Processing-in-Memory (PIM)~\cite{kwon2022system, he2020newton, Vetter.2023, Ausavarungnirun.2020} has been  proposed to accelerate LLM by exploiting high internal memory bandwidth~\cite{choi2024attacc, heo2024neupims, cent}. NeuPIMs~\cite{heo2024neupims} proposes an NPU+PIM heterogeneous system where NPU is leveraged for compute-intensive kernels (i.e., GEMM) while PIM is leveraged for memory-bound kernels (i.e., GEMV) to accelerate LLM inference. PIM-only systems (e.g., CENT~\cite{cent}) have also been proposed as a multi-PIM node system to handle large LLMs via CXL-based memory expansion. While these prior works accelerate LLM inference with PIM, they primarily focus on relatively small input context lengths (e.g., 4K\footnote[1]{CENT does provide ablation study where they scale to 32K but the benefits of PIM decrease as sequence length increases. More importantly, this work identifies PIM inefficiency as the context length increases and maximizes overall performance from PIM.}). In this work, we revisit PIM for LLM inference but focus on long-context LLM (up to 1M tokens) and identify how PIM inefficiency becomes problematic as the Attention layers in decoding become a greater bottleneck with longer context. In long-context scenarios, PIM inefficiency arises from workload imbalance that leaves channels underutilized and from a fixed write–compute–read pipeline that creates I/O bottlenecks and stalls processing units. KV cache remains problematic in terms of PIM memory capacity but \emph{static} memory allocation based on maximum context length becomes problematic when context length can vary significantly across workloads~\cite{longchat2023,bai2024longbench,lveval}.

To address these limitations, we propose \arch, a PIM \emph{orchestrator} that enables efficient data mapping and movement to improve PIM utilization through three co-designed techniques. First, unlike the conventional head-first or batch-dimension partitioning approach, \textit{Token-Centric PIM Partitioning (TCP)} distributes tokens across all channels within a single PIM module to ensure high utilization and load-balancing regardless of batch size. Second, to mitigate the PIM I/O bottleneck, \textit{Dynamic PIM Command Scheduling (DCS)} issues commands based on real-time data dependencies to maximally overlap computation and data movement—a capability absent in static PIM schedulers. Finally, \textit{Dynamic PIM Access (DPA)} overcomes the limitations of static memory management; by embedding loop bounds and operand modifications into the command stream, it enables dynamic KV cache allocation to improve memory utilization. As a result, PIMphony achieves higher efficiency by better exploiting the high internal bandwidth of each PIM channel by balancing workloads, continuously feeding data to its MAC units, and allocating memory on-demand.

\begin{figure}[t]
    \centering
    \includegraphics[width=0.9\linewidth]{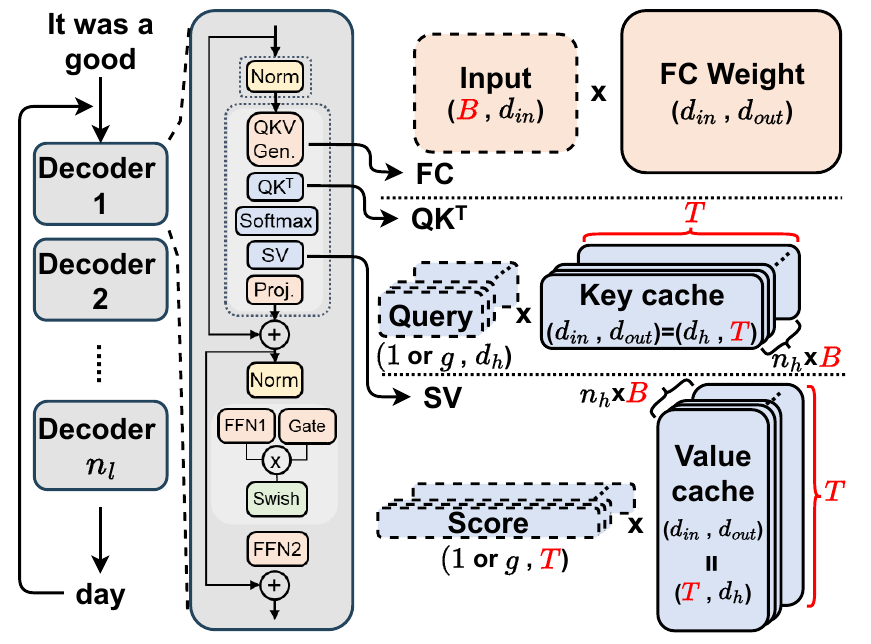}
    \vspace{-5pt}
    \caption{Decoding Computation for Long-Context LLM ($g$: group size of GQA~\cite{ainslie2023gqa})}
    \label{fig:decoding}
\end{figure}

\begin{table}[t]
    \centering
    \caption{LLM specification and the context window (CW).}
    \begin{adjustbox}{width=\linewidth}
    \begin{tabular}{c|c|c|c|c|c|c|c}
        \hline
        \makecell{Model} & $n_l$ & $n_h$ & $d_h$ & $d_{in/out}$ & GQA & \makecell{Reference} & CW \\ \hline \hline
        \multirow{2}{*}{LLM-7B} & \multirow{2}{*}{32} & \multirow{2}{*}{32} & \multirow{2}{*}{128} & 4K-12K & \(\times\) & QWEN1.5-7B \cite{bai2023qwen} & 32K \\ \cline{5-8}
         & & & & 4K-12K & \checkmark ($g=4$) & Llama3.1-8B \cite{dubey2024llama3herdmodels} & 128K \\ \hline
        \multirow{2}{*}{LLM-72B} & \multirow{2}{*}{80} & \multirow{2}{*}{64} & \multirow{2}{*}{128} & 8K-24K & \(\times\) & Qwen1.5-72B \cite{bai2023qwen} & 32K \\ \cline{5-8}
         & & & & 8K-24K & \checkmark ($g=8$) & Llama3.1-70B \cite{dubey2024llama3herdmodels} & 128K \\ \hline
    \end{tabular}
    \end{adjustbox}
    \label{tab:LLMs}
    \vspace{-10pt}
\end{table}

We implement \arch by extending an MLIR-based compiler and runtime to generate the PIM commands for both token-level partitioning and dynamic KV cache allocation. We developed custom compiler passes that detect transformer decoder patterns, generate PIM commands with dynamic row/column indices, and handle the token-parallel mapping within each module. We also modified a cycle-accurate DRAM simulator to model PIM I/O output buffering and apply \arch to both a heterogeneous architecture (i.e., NeuPIMs~\cite{heo2024neupims}) and a PIM-only architecture (i.e., CENT~\cite{cent}). Across representative long-context LLM workloads, our evaluations show up to 11.3$\times$ improvement in performance compared to state-of-the-art PIM accelerators for LLM inference. 

\begin{figure}[t]
    \centering
    \includegraphics[width=1\linewidth]{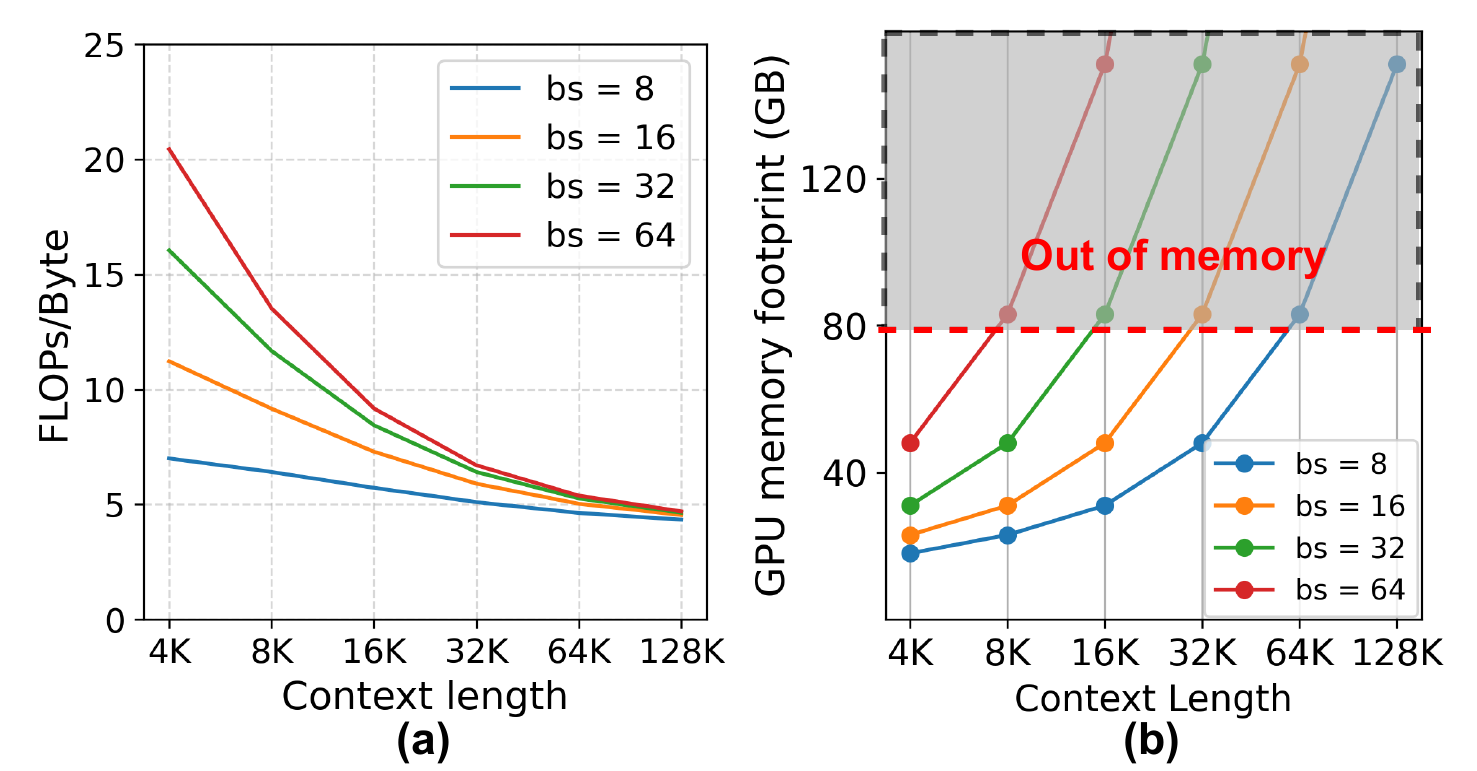}
    \vspace{-20pt}
    \caption{Characteristics of long-context LLM decoding on LLM-7B (w/ GQA). (a) Compute intensity (FLOPs/Byte) decreases with context length. (b) GPU memory footprint grows with both context length and batch size; the dashed line marks the A100-80GB capacity.}
    \label{fig:Characteristic-LongContextLL}
\end{figure}

\begin{table}[!t]
    \centering
    \caption{Statistics of input context length. 
    }
    \begin{adjustbox}{width=0.9\linewidth}
    \begin{tabular}{c|c|c|c|c}
    \hline
    \multirow{2}{*}{Statistic} & \multicolumn{2}{c|}{\textbf{LongBench}~\cite{bai2024longbench}} & \multicolumn{2}{c}{\textbf{LV-Eval}~\cite{lveval}} \\
    \cline{2-5}
     & QMSum & Musique & multifieldqa & Loogle-SD\\
    \hline \hline
    mean & 13,966  & 16,362 & 60,780 & 50,693\\
    std  & 6,182  & 1,651  & 31,025 & 26,506\\
    max  & 30,456 & 17,917 & 119,480 & 109,221\\
    min  & 2,651   & 6,820  & 20,333 & 13,347\\
    \hline
    \end{tabular}
    \end{adjustbox}
    \label{tab:longbench_statics}
    \vspace{-10pt}
\end{table}

\section{Background/Motivation}

\subsection{Long-Context LLM Inference/Decoding}
\label{subsec:background_lllm}

Long-context LLMs are built on the Transformer decoder architecture~\cite{vaswani2017attention}, which, as shown in Fig.~\ref{fig:decoding}, consists of $n_l$ layers, each containing a Multi-Head Attention (MHA) module and a Feed-Forward Network (FFN). Within each of the $n_h$ attention heads, per-head feature dimension ($d_h$), matrix dimension ($d_{in}$, $d_{out}$) for weight parameters, Query/Key/Value vectors (${Q,K,V} \in \mathbb{R}^{d_h}$) are generated, and the K/V vectors are appended to a KV cache for all $T$ tokens. To improve efficiency, some models employ architectural variants like Grouped-Query Attention (GQA)~\cite{ainslie2023gqa}, where multiple query heads share a single set of Key and Value vectors per group. The subsequent Attention operations, $QK^T$ and $SV$, access this cache. Representative LLM configurations, including GQA variants, are detailed in Table~\ref{tab:LLMs}, and benchmark characteristics are outlined in Table~\ref{tab:longbench_statics}.

Our analysis of long-context LLM workloads reveals memory bandwidth and capacity bottlenecks from Attention during decoding. As context length grows, compute intensity (OPs/Byte) drops sharply (Fig.~\ref{fig:Characteristic-LongContextLL}(a)) as computation shifts from compute-intensive matrix-matrix (GEMM) to memory-bound matrix-vector (GEMV) operations for Attention—making performance heavily dependent on raw memory bandwidth. This bandwidth pressure is compounded by capacity demand of the KV cache. Because the KV cache increases proportionally with context length and batch size, it dominates the growth of overall memory requirements and thus imposes significant capacity pressure in long-context LLM inference (Fig.~\ref{fig:Characteristic-LongContextLL}(b)).
In this work, we exploit a multi-node PIM system to provide both the required memory capacity and high internal memory bandwidth.

\begin{figure}[t]
    \centering
    \includegraphics[width=0.8\linewidth]{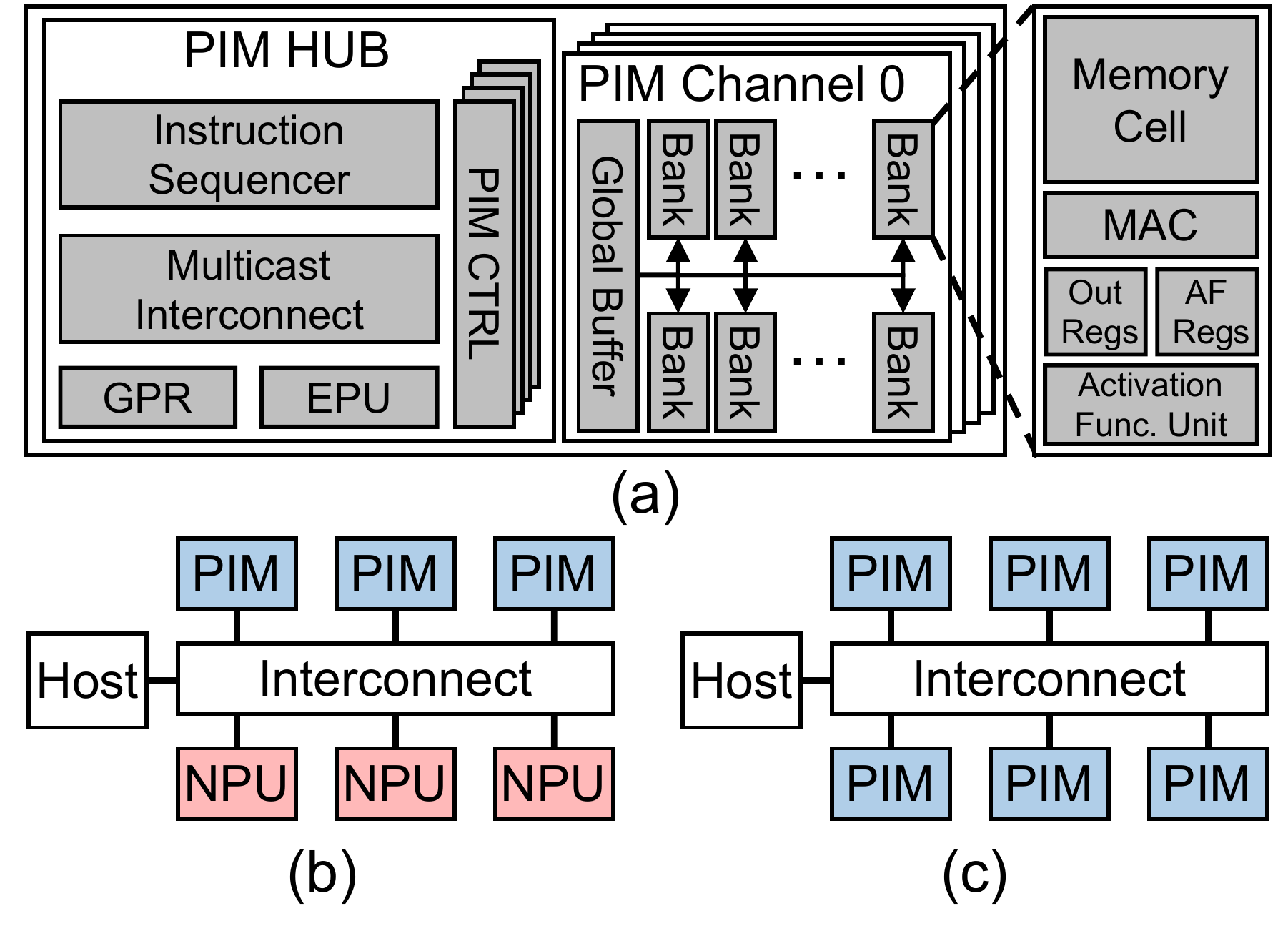}
    \vspace{-10pt}
    \caption{Overview of PIM module/node configuration. (a): PIM module architecture. (b) and (c): PIM node configuration - heterogeneous xPU+PIM and PIM-only.}
    \label{fig:System_Overview}
    \vspace{-10pt}
\end{figure}

\subsection{PIM Architecture and Instruction Execution}
\label{subsec:2.B}

DRAM-based Processing-in-Memory (PIM) systems~\cite{AiM_JSSC,AiM_ISSCC} accelerate memory-bound GEMV operations by integrating computation directly into memory. As detailed in Fig.~\ref{fig:System_Overview}(a), a PIM module integrates vector MAC units within each DRAM bank, a shared Global Buffer (GBuf) for inputs, and Output Registers (OutRegs) for results, all orchestrated by a PIM Controller. 
An Activation Function Unit, with an associated Activation Function Registers, is also included to support non-linear activation functions via Look-Up Table (LUT) approximations.
An Extra Processing Unit (EPU) performs auxiliary operations (e.g., Softmax), while a General-Purpose Register file (GPR) stores inputs/outputs (or intermediate results) of the computation. 
In operation, the PIM HUB receives compiler-generated PIM instructions. 
The Instruction Sequencer expands each instruction by unrolling its specified repetitions and forwards the resulting instruction stream to the Multicast Interconnect. 
The Multicast Interconnect then decodes these instructions into channel-specific PIM commands, multicasts them to the target channels, and routes the associated data to the corresponding PIM Controllers.

Each PIM instruction carries a set of arguments that determine how it is expanded and dispatched as PIM commands (Table~\ref{tab:PiM-instructions}). Specifically, \textit{Ch-mask} specifies the target PIM channels, \textit{Op-size} determines repetition count, and \textit{GPR-addr} provides the base access address when the instruction involves data movement between the GPR and the PIM (e.g., \texttt{WR-INP} or \texttt{RD-OUT}). The Instruction Sequencer uses \textit{Op-size} to unroll a single instruction into a sequence of repeated instructions, which are subsequently decoded by the Multicast Interconnect into channel-specific PIM commands accessing consecutive addresses (e.g., GPR-addr, GBuf-Idx, or column addresses).

\begin{table}[t]
    \centering
    \caption{PIM instructions for LLM inference. \textit{Ch-mask}, \textit{Op-size}, \textit{GPR-addr} guide the decoding of each instruction into channel-specific PIM commands, while \textit{GBuf-Idx}, \textit{Out-Idx}, \textit{Row/Col} are used during PIM channel operation.}
    \begingroup
    \renewcommand{\arraystretch}{1.2}
    \begin{adjustbox}{width=\linewidth}
    \begin{tabular}{c|c|c}
    \hline 
    Instruction & Description & Arguments \\
    \hline \hline
    \texttt{WR-INP} & Copy Input from GPR to GBuf & \makecell[l]{Ch-mask Op-size GPR-addr\\GBuf-Idx} \\
    \hline
    \texttt{MAC} & Dot-Product on a DRAM row & \makecell[l]{Ch-mask Op-size\\GBuf-Idx Row/Col Out-Idx} \\
    \hline
    \texttt{RD-OUT} & Copy Output from OutReg to GPR & \makecell[l]{Ch-mask Op-size GPR-addr\\ Out-Idx} \\
    \hline
    \end{tabular}
    \end{adjustbox}
    \endgroup
    \label{tab:PiM-instructions}
    \vspace{-10pt}
\end{table}

\subsection{Multi-Node DRAM-PIM System}
To accommodate large models, multiple nodes can be deployed in heterogeneous (xPU+PIM) or PIM-only configurations (Fig.~\ref{fig:System_Overview}(b,c)), leveraging model parallelism techniques such as Tensor Parallelism (TP)~\cite{Shoeybi.2019} and Pipeline Parallelism (PP)~\cite{huang2019gpipe}. In TP, model parameters are partitioned across PIM modules, and parallel computation is performed on different tensor shards, with synchronization required to aggregate partial results. In contrast, PP partitions the model into transformer layers that are executed across PIM modules, allowing different micro-batches to be processed concurrently in a pipelined manner. Both approaches enable large transformer models to be mapped across multiple PIM modules by exposing parallelism at different granularities.

\subsection{Challenges}
\label{subsec:challenges}

\begin{figure}[!t]
    \centering
    \includegraphics[width=1\linewidth]{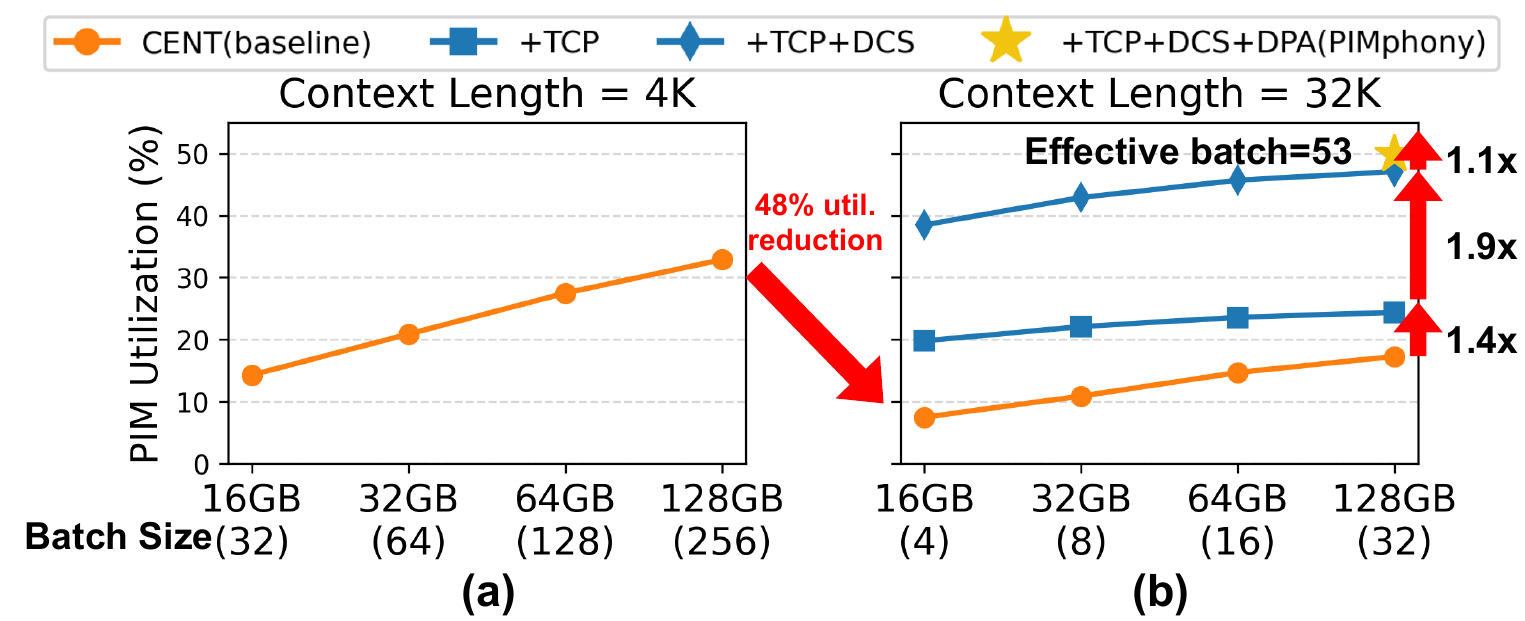}
    \vspace{-15pt}
    \caption{
    PIM utilization under (a) short(4K) and (b) long(32K) contexts using CENT~\cite{cent} and PIMphony on LLM-7B-32K-GQA. Batch size scales inversely with context length due to the capacity constraint.}
    \label{fig:challenge_util}
    \vspace{-15pt}
\end{figure}

Although PIM systems offload memory-bound Attention operations, their efficiency drops sharply in long-context scenarios. We identify three fundamental limitations that create a critical performance bottleneck:

\begin{enumerate}
    \item \textbf{Channel Underutilization.} Prior PIM systems partition the KV cache by the head and batch dimensions to parallelize Attention across PIM channels~\cite{cent,choi2024attacc,heo2024neupims}. However, this strategy is highly inefficient for long contexts. It leads to severe channel underutilization, either from workload imbalance when requests have different context lengths or from insufficient sub-batch sizes to keep all pipeline stages full. 
    \item \textbf{I/O Bottleneck.} Each dot-product in PIM follows a fixed command pipeline of \texttt{WR-INP}~$\rightarrow$~\texttt{MAC}~$\rightarrow$~\texttt{RD-OUT} (Table~\ref{tab:PiM-instructions}), but limited I/O buffers—2KB input per channel and 4-byte output per bank~\cite{AiM_ISSCC}—frequently stall the MAC units. This issue is amplified in Attention layers, where the KV cache has a small feature dimension ($d_{h}$; Fig.~\ref{fig:decoding}), limiting input/output reuse and increasing buffer turnover. Critically, static command scheduling fails to overlap data movement with computation, resulting in low MAC utilization even under optimal partitioning.
    \item \textbf{Static Memory Management.} To handle variable context lengths, existing PIM systems pre-allocate a KV cache sized for the maximum context $T_{max}$, bounding the batch size to a worst-case capacity. Given that real-world workloads have diverse token lengths (Table II), this static approach leads to severe memory inefficiency, with an observed average capacity utilization of only 36.2\% (See Sec.~\ref{SubSec:Ablation}). The root cause is that PIM commands embed fixed physical addresses, making it impossible to repurpose these large, unused memory regions at runtime.
    
\end{enumerate}
Taken together, the challenges of channel underutilization, I/O bottlenecks, and static memory waste show that conventional PIM systems are fundamentally underutilized for long-context LLM inference. This inefficiency is substantial; our analysis reveals that MAC unit utilization drops by 48\% at a 32K context length (Fig.~\ref{fig:challenge_util}). Such systemic limitations cannot be solved with incremental optimizations, necessitating a holistic redesign of the PIM software and hardware stack—motivating the novel orchestrator we propose.

\section{\arch Overview}
\label{sec:PIMphony_overview}
\begin{figure}[t]
    \centering
    \includegraphics[width=\linewidth]{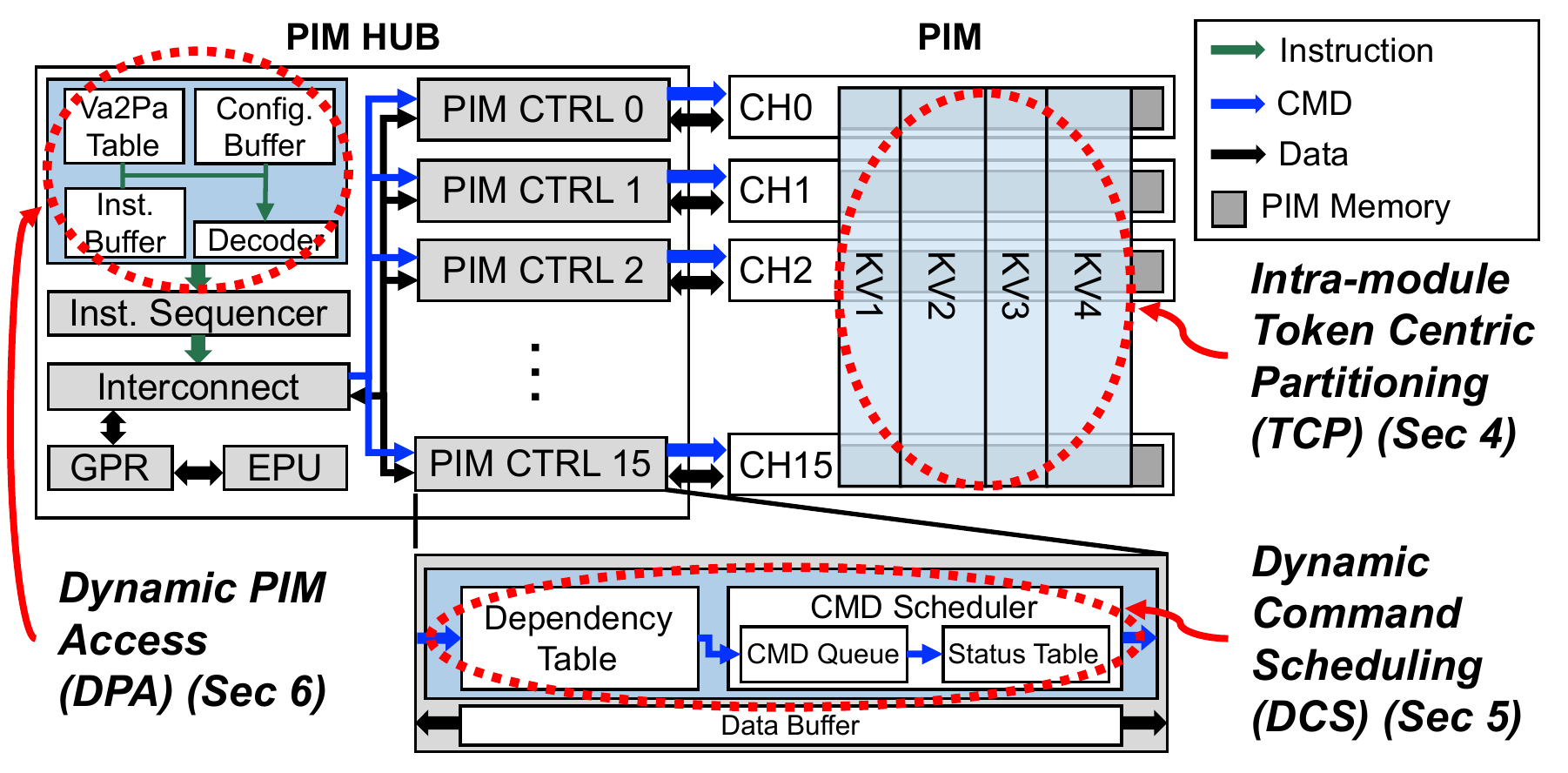}
    \vspace{-10pt}
    \caption{High-level overview of \arch with the three main components highlighted.}
    \label{fig:PIMphony_Overview}
    \vspace{-15pt}
\end{figure}

We propose PIMphony, a PIM \textit{orchestrator} that systematically resolves these inefficiencies to enable high performance for long-context LLM inference. PIMphony improves PIM utilization—a critical yet previously underexplored bottleneck—through three co-designed techniques (Fig.~\ref{fig:PIMphony_Overview}). 

\textbf{Token-Centric PIM Partitioning (TCP).}  
To combat channel underutilization, PIMphony introduces TCP. Unlike conventional methods that rely on the volatile batch dimension, TCP reorients parallelism along the plentiful \textit{token} axis. By distributing token-level work across all channels, TCP decouples performance from batch size, mitigating imbalance and ensuring consistently high channel utilization.

\textbf{Dynamic PIM Command Scheduling (DCS).}  
To eliminate the I/O bottleneck, PIMphony employs DCS, which enhances the PIM controller with the ability to issue commands out-of-order based on real-time data dependencies. This dynamic approach, impossible for rigid static schedulers, effectively hides I/O latency by overlapping data movement and computation, thus maximizing MAC pipeline throughput.

\textbf{Dynamic Memory Management with Dynamic PIM Access (DPA).} 
To address wasteful static memory allocation, PIMphony features DPA, which includes a novel on-module dispatcher, which effectively acts as a lightweight, pseudo-Memory Management Unit (MMU) for the PIM. This dispatcher enables runtime virtual-to-physical address translation, allowing for dynamic, on-demand memory allocation for the KV cache. This breaks free from the static pre-allocation model based on maximum context length and improves effective memory capacity utilization.

\begin{figure*}[t]
    \centering
    \includegraphics[width=1\textwidth]{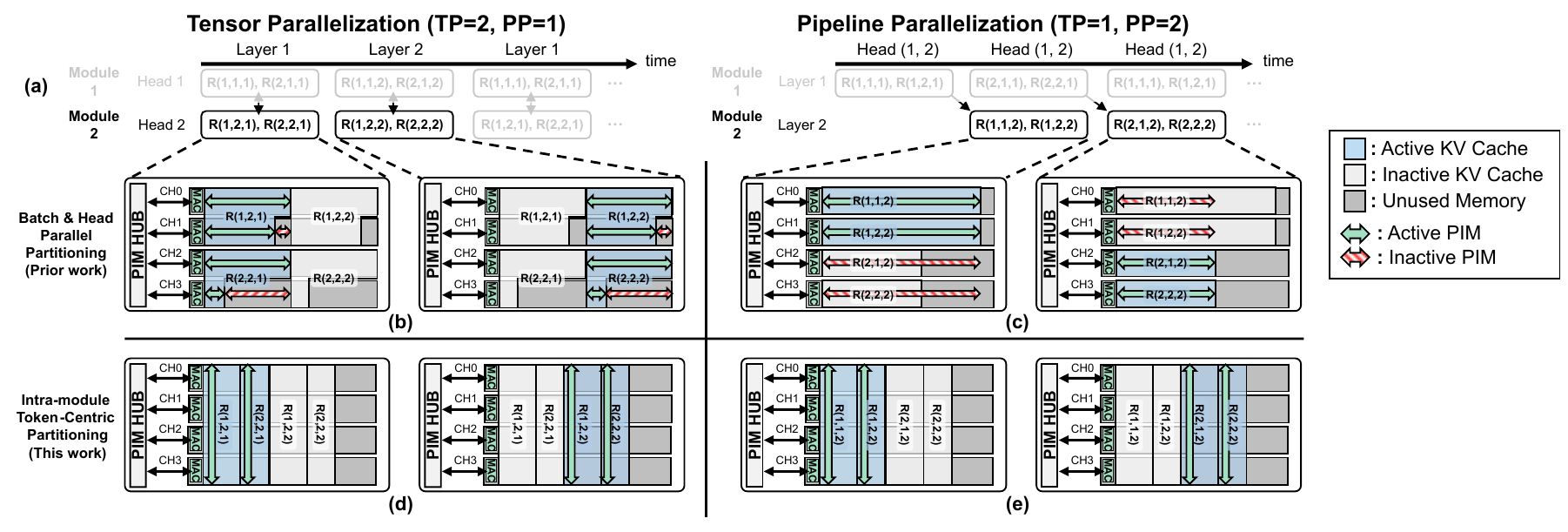}
    \vspace{-20pt}
    \caption{Comparison of KV cache partitioning strategies across PIM channels for tensor parallelism (TP) (b,d) and pipeline parallelism (PP) (c,e). Prior approach~\cite{cent,choi2024attacc,heo2024neupims} (Head/Batch-First Partitioning (HFP)) is shown in (b,c) while the proposed Token-Centric Partitioning (TCP) is shown in (d,e). The example highlights the inefficient utilization of HFP across the PIM channels, as not all channels can be utilized simultaneously. For clarity, only one PIM module with four channels is shown. R($r$, $h$, $l$) indicates Request $r$, Head $h$, and Layer $l$.}
    \vspace{-10pt}
    \label{fig:Channel_Parallelism}
\end{figure*}

Together, TCP, DCS, and DPA form a cohesive orchestration framework that addresses PIM inefficiencies across parallelism, command scheduling, and memory management.
Fig.~\ref{fig:challenge_util}(b) provides a preview of their cumulative impact on PIM utilization.
Building on this architectural overview, the following sections describe each technique in detail, explaining how they jointly enable efficient long-context LLM inference.

\section{Token-Centric PIM Partitioning}
\label{sec:TCPP}
To understand the limitations of existing PIM workload partitioning strategies for long-context LLM inference, we first examine a simplified example that illustrates their impact on channel utilization. Fig.~\ref{fig:Channel_Parallelism}(a) visualizes a Transformer example workload with two attention heads, two layers, and a batch size of two, distributed across two PIM modules, each containing four independently operating channels. Computation tasks are labeled with the notation $R(r,h,l)$, where $r$ denotes the request ID, $h$ the head index, and $l$ the layer index, while active and inactive KV cache regions are shown in blue and shade, respectively. This example reflects long-context conditions, where a request typically consumes nearly the entire memory capacity of a single PIM channel, limiting each channel to serve only one request at a time. This example allows us to compare the effects of tensor parallelism (TP) and pipeline parallelism (PP) on channel activity, contrasting conventional PIM partitioning with our proposed token-centric PIM partitioning (TCP) approach.

\subsection{Channel Underutilization Issue}
\label{subsec:chunderutil}
Modern PIM-based LLM accelerators~\cite{cent,choi2024attacc,heo2024neupims} suffer severe channel underutilization in long-context inference because existing KV cache partitioning schemes do not scale with growing context length. To distribute workload across channels within a PIM module, prior systems commonly adopt Head-First Partitioning (HFP), which assigns head–batch pairs to individual PIM channels for concurrent execution.
HFP implicitly assumes the availability of sufficient head–batch parallelism to populate all channels. However, as context length increases, each request’s KV cache footprint expands while the number of simultaneous requests (i.e., batch size) shrinks. In the extreme, a single long-context request can occupy an entire channel, drastically reducing the number of available head–batch tiles.
As a result, HFP fails to sustain high channel utilization in long-context settings, leaving many channel-level MAC units idle despite abundant token-level parallelism. In the following, we analyze how this limitation of HFP manifests under tensor and pipeline parallelism.

\subsection{Analysis of HFP under Tensor and Pipeline Parallelism}
Tensor parallelism (TP) and pipeline parallelism (PP) organize LLM execution across PIM modules at different granularities—TP partitions attention heads across modules, while PP distributes consecutive layers.
Under both schemes, however, the intra-module channel allocation remains governed by head-first partitioning (HFP), which ultimately determines how workloads are mapped to individual PIM channels.

\textbf{Load Imbalance from HFP under TP.}
Under TP, HFP leads to load imbalance when requests have different token lengths, leading to varied execution times across channels.
Channels assigned to shorter sequences finish their computations earlier and must wait for those processing longer sequences, limiting the overall throughput to that of the slowest channel and reducing effective parallelism.
For example, in Fig.~\ref{fig:Channel_Parallelism}(b), Channel 2 and 3 are assigned to Request~2 with a shorter token length and thus performs less computation, becoming idle while other channels continue execution.
Although short-context systems can mitigate such imbalance by assigning multiple requests per channel for load balancing (e.g., NeuPIMs~\cite{heo2024neupims}), this approach becomes infeasible in long-context inference, where a single request can fully occupy a channel’s memory capacity.

\textbf{Sparse Channel Activation from HFP under PP.} Under PP, HFP activates only the subset of channels associated with the request assigned to each pipeline stage.
In long-context scenarios, where a single request can occupy an entire channel, this results in sparsely populated stages with many idle channels. As illustrated in Fig.~\ref{fig:Channel_Parallelism}(c), only a fraction of the available channels are active at any given time. This stage-level idling, compounded by pipeline bubbles that form when subsequent stages are empty, results in persistent underutilization—a problem especially pronounced with the limited batch sizes of long-context workloads.

\subsection{Intra-Module Token-Centric Partitioning}
\label{subsec:TCP.B}
To overcome the channel underutilization inherent in HFP, we introduce Token-Centric PIM Partitioning (TCP). 
TCP partitions the token dimension of a single head across all available PIM channels, enabling token-level parallelism within each PIM module.
In $QK^T$ operations, each channel processes a distinct segment of tokens concurrently, enabling parallel computation across the module.
In $SV$ operations, each channel performs token-wise partial reduction over its assigned tokens, and the partial results are reduced through the shared PIM HUB and GPR to produce the final output.
For example, in Fig.~\ref{fig:Channel_Parallelism}, assume each channel contains 16 banks, with a total token length of 16K and a head dimension of 32. Under this configuration, $QK^T$ assigns 4K tokens to each channel for parallel computation, while $SV$ assigns 2K tokens per channel and performs a single global reduction across channels through the PIM HUB to generate the final result.

\begin{figure*}[t]
    \centering
    \includegraphics[width=1\textwidth]{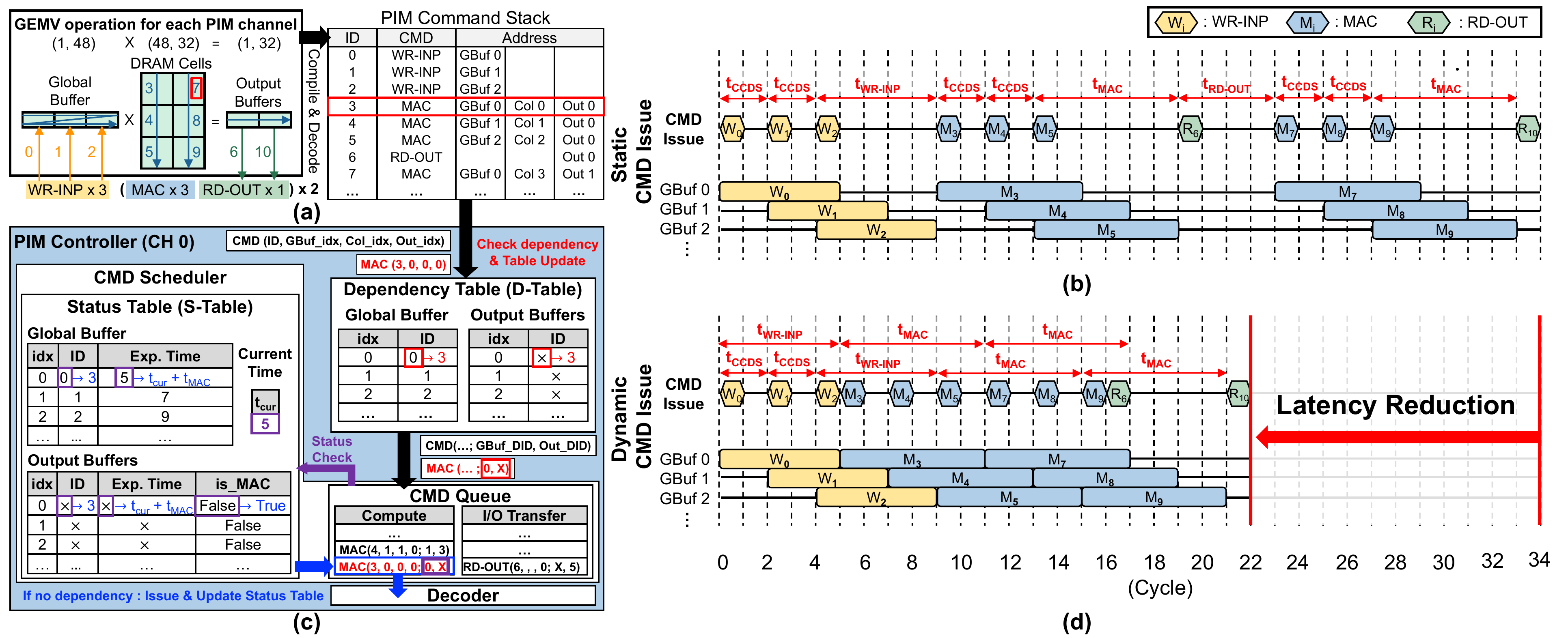}
    \vspace{-15pt}
    \caption{Dynamic PIM Command Scheduling. (a) A GEMV operation example and its command stack,  (b) timing diagram for baseline PIM command schedule,  (c) detailed Dynamic PIM Command Scheduling (DCS) example within the PIM controller, and (d) resulting timing diagram after DCS where \texttt{MAC} instructions can be executed in advance. In (b)\&(d), Each GB 0/1/2 denotes an entry of a Global Buffer.}
    \vspace{-10pt}
    \label{fig:pim-cmd-schedule}
\end{figure*}

TCP enables full channel activation in long-context inference, 
where the token sequence length is sufficiently large to keep all processing units active. As illustrated in Fig.~\ref{fig:Channel_Parallelism}(d) and (e), TCP distributes token computations across channels, ensuring full channel activation. Under TP, TCP mitigates token-length imbalance across requests that previously caused idle channels, while under PP, TCP allows all channels to participate concurrently regardless of the active pipeline stage. Under a commercial PIM module~\cite{AiMX-White-Paper} configuration with 16 channels and 16 banks per channel, full channel activation is achieved once the token length exceeds 256 for QKT and 32 for SV.

Under TCP, each PIM channel produces partial outputs that must be combined within the module to form a complete result.
This aggregation step operates at the inter-channel level, where outputs from different channels are gathered through the general-purpose registers (GPR) at the PIM HUB and finalized by the Extra Processing Unit (EPU) (Fig.~\ref{fig:System_Overview}(a)).
For $QK^T$, the aggregation after matrix multiplication involves only concatenation during the subsequent Softmax in the EPU, incurring no measurable latency. Meanwhile, SV performs a single inter-channel reduction per module, whose cost is minimal—below 0.2\% of total attention latency for an LLM-7B with 16K tokens. Because TCP partitions tokens only within a module, it avoids inter-module synchronization, keeping aggregation and synchronization overhead negligible.

\section{Dynamic PIM Command Scheduling}
\label{sec:double-buf}

\subsection{PIM Command Execution Overview}
\label{subsec:DCS_command}
Modern PIM architectures follow a command-driven execution model where primitive operations—\texttt{WR-INP}, \texttt{MAC}, and \texttt{RD-OUT}—are issued sequentially to hardware units (see Table~\ref{tab:PiM-instructions}). 
\texttt{WR-INP} writes a 32B tile to a Global Buffer (GBuf) entry, \texttt{MAC} reads that entry for multiplication and accumulates results into per-bank OutRegs\, and \texttt{RD-OUT} drains a 2B result from all 16 banks concurrently (32B in total). 
These primitives are composed into a command stack to perform computation. For example, the FP16 GEMV in Fig.~\ref{fig:pim-cmd-schedule}(a) is executed by streaming the input tiles via \texttt{WR-INP}, accumulating partial dot products with \texttt{MAC}, and retrieving the output tiles with \texttt{RD-OUT}. Due to the pipelined operation of the data bus, transferring 32B tiles has a minimum command-to-command interval ($t_{CCDS}$). As illustrated in Fig.~\ref{fig:pim-cmd-schedule}(b), successive \texttt{WR-INP}s ($W_0$, $W_1$, $W_2$) are issued $t_{CCDS}$ cycles apart, pipelining the streaming of input tiles.

Conventional PIM controllers~\cite{AiMX-Hotchips-2022,AiMX-White-Paper} execute commands using \textit{static scheduling}:
the controller issues commands strictly in-order in a fixed \texttt{WR-INP} $\rightarrow$ \texttt{MAC} $\rightarrow$ \texttt{RD-OUT} pattern (as generated from the instruction stream). To avoid data hazards, it decides issue by enforcing time gaps derived from fixed command execution times ($t_{\text{WR-INP}}$, $t_{\text{MAC}}$, $t_{\text{RD-OUT}}$) between commands. For example, a \texttt{MAC} must wait at least $t_{\text{WR-INP}}$ after the preceding \texttt{WR-INP} to ensure the input tile is fully written into the GBuf. As shown in Fig.~\ref{fig:pim-cmd-schedule}(b), this approach needlessly serializes operations, causing pipeline stalls even when no true data dependency exists between commands (e.g., the input write $W_2$ and the computation $M_3$, or the output read $R_6$ and computation $M_7$).

\subsection{I/O Bottleneck Analysis}
\label{subsec:io-bottleneck-analysis}

Attention layers inherently incur frequent I/O transfers (\texttt{WR-INP}, \texttt{RD-OUT}) due to their significantly lower data reuse compared to fully-connected (FC) layers.
This effect is particularly pronounced in Attention's core operations: in $QK^T$, a small input dimension ($d_{in}$) reduces output reuse, while in $SV$, a small output dimension ($d_{out}$) limits input reuse. In both cases, the reduced data reuse necessitates more frequent I/O transfers. Moreover, under conventional PIM controllers, these frequent transfers are executed with \textit{static scheduling}, which can further amplify the resulting performance bottleneck.

\begin{figure}[t]
    \centering
    \includegraphics[width=1\linewidth]{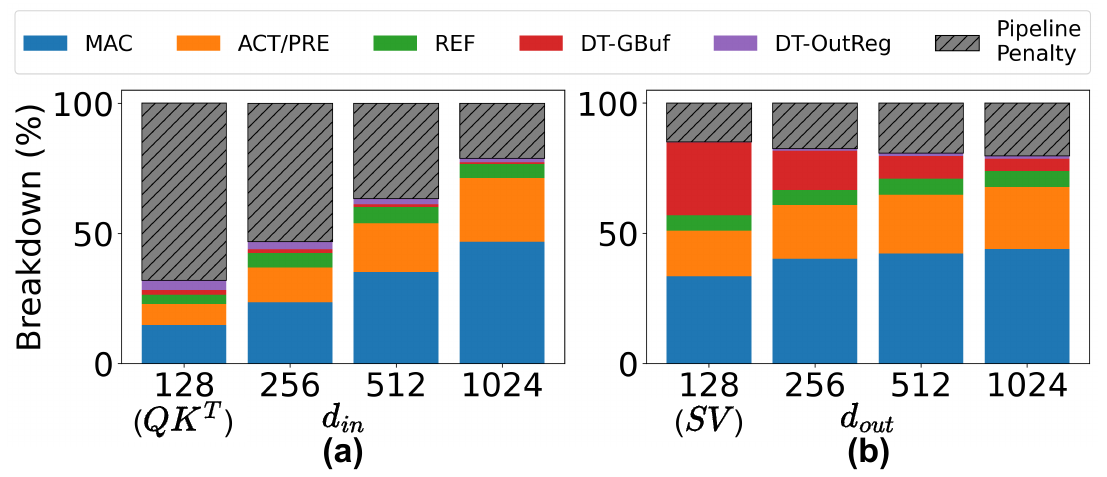}
    \vspace{-15pt}
     \caption{Latency breakdown across matrix dimensions. MAC is computation time; ACT/PRE and REF are DRAM activation/precharge and refresh time; DT-GBuf and DT-OutReg represent I/O transfers time; Pipeline Penalty captures cumulative stalls across PIM commands.}
    \label{fig:IO_bottleneck}
    \vspace{-10pt}
\end{figure}

\textit{Static scheduling} enforces a fixed command ordering and timing without tracking per-entry data dependencies across GBuf and OutRegs. 
Consequently, it fails to overlap data transfer and computation, serializing them even when data hazards are cleared. This conservative scheduling blocks MAC execution even when resources are available, leading to substantial pipeline penalties and idle cycles. 
Fig.~\ref{fig:IO_bottleneck} quantifies the impact of this inefficiency, showing that as matrix dimensions ($d_{in}$, $d_{out}$) decrease, pipeline stall and I/O transfer time become dominant. In particular, for small dimensions typical of Attention (128, corresponding to a head dimension), MAC utilization drops sharply to 14.7\%. This trend indicates that frequent I/O further exacerbates the limitations of static scheduling.
Since frequent I/O is unavoidable in Attention workloads, this fundamental limitation of \textit{static scheduling} motivates the need for a more flexible, command-level, dependency-aware Dynamic Command Scheduling (DCS).

\subsection{Dynamic PIM Command Scheduling}
\label{subsec:DCS.C}
\textbf{I/O-aware Buffering.} 
To mitigate the bottleneck from frequent I/O transfer, we introduce I/O-aware buffering that decouples I/O transfers from computation, allowing the controller to issue independent PIM commands more aggressively and keep the pipeline busy. For this, we repurpose the existing multi-entry GBuf as a dedicated input buffer and expand the Output Registers into larger Output Buffers (OBuf) for output staging. Both of these buffers are implemented as dual-port memory, enabling concurrent accesses: while \texttt{MAC} consumes the current inputs and accumulates partial sums, the controller can prefetch the next inputs into GBuf or drain completed outputs from OBuf, increasing I/O--compute overlap.

\textbf{Dynamic Command Scheduling.} 
We extend the PIM controller with a dynamic scheduling mechanism that enforces inter-command timing constraints only when true data dependencies exist between I/O transfer and \texttt{MAC} operations. To support this, we introduce two hardware structures—the Dependency Table (D-Table) and the Status Table (S-Table)—as illustrated in Fig.~\ref{fig:pim-cmd-schedule}(c).

Each command is assigned a unique identifier (ID) and maintains a Dependency ID (DID), which specifies the ID of the command that must complete before the current command can be safely executed. The D-Table enables this dependency assignment by recording, for each GBuf and OBuf entry, the most recent command that accessed it. When a new command arrives, the controller consults the corresponding D-Table entry, assigns the recorded command ID as the new command’s DID, and updates the D-Table with the new command ID.

The S-Table maintains the execution status to determine whether a dependency has been resolved. For each GBuf and OBuf entry, it records the ID of the most recent accessing command along with the cycle at which that access completes (i.e., an expiration timestamp). In addition, the S-Table maintains an \textit{is-MAC} flag for OBuf entries to identify consecutive \texttt{MAC} operations targeting the same buffer location, enabling pipelined execution without unnecessary stalls.

Together, these structures govern command scheduling and dispatch. After dependency assignment via the D-Table, the controller classifies each command and enqueues it into either the I/O transfer queue or the compute queue, allowing out-of-order execution between I/O transfer commands and \texttt{MAC} commands while preserving in-order execution within each queue. For each command considered for execution, the controller consults the S-Table to determine its execution readiness. A command is dispatched only (i) if the S-Table entry corresponding to its accessed buffer reports a matching command ID to the command’s DID and (ii) the current cycle $t_{\text{cur}}$ exceeds the recorded expiration time, ensuring that the predecessor command has completed. When these conditions are satisfied, the controller issues the command and updates the S-Table to reflect the new access and its completion time. 
For consecutive \texttt{MAC} operations accessing the same OBuf entry, the \textit{is-MAC} flag allows the controller to bypass the $t_{\text{MAC}}$ gap and issue them at the minimum $t_{CCDS}$ interval.

\begin{figure}[t]
    \centering
    \includegraphics[width=1\linewidth]{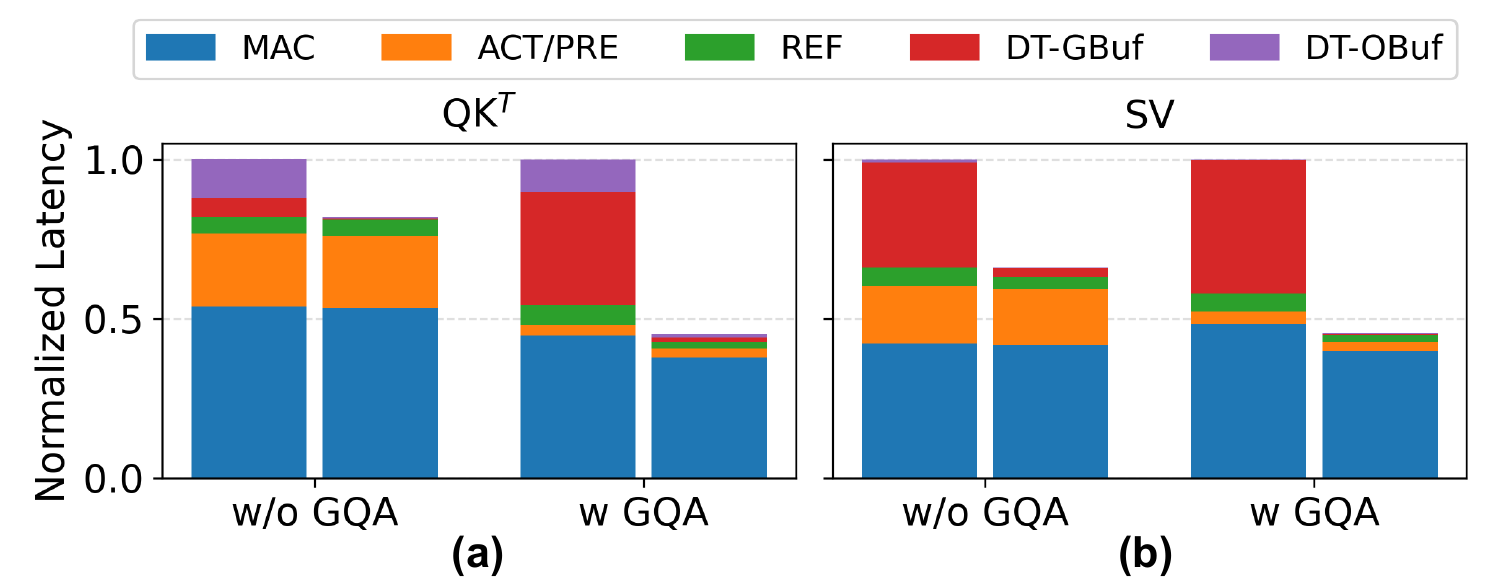}
    \vspace{-20pt}
    \caption{Latency breakdown of PIM command execution for LLM-72B Attention. (a) $QK^T$, (b) $SV$. Each bar compares execution without (left) and with DCS (right). 
    Both employ the \emph{row-reuse mapping}}
    \label{fig:GQA_breakdown}
    \vspace{-10pt}
\end{figure}

\textbf{PIM Controller Example.} 
Fig.~\ref{fig:pim-cmd-schedule}(c) illustrates an example of the PIM controller operation. When command $M_3$ arrives, the controller consults the D-Table and identifies that command $W_0$ was the most recent command to access GBuf entry 0, which $M_3$ targets. Accordingly, $M_3$ is assigned a GBuf Dependency ID of 0. This dependency association indicates that $M_3$ must wait for $W_0$ to complete before accessing the GBuf entry.
To determine whether this dependency has been resolved, the controller examines the S-Table entry corresponding to GBuf index 0. Issuance of $M_3$ is permitted only if the command ID recorded in the S-Table matches $M_3$’s GBuf-DID and the current cycle $t_{cur}$ exceeds the recorded expiration timestamp. This check ensures that $M_3$ is issued only after $W_0$ has completed its access.

\begin{figure*}[t!]
    \begin{minipage}[b]{0.61\textwidth}
        \centering
        \includegraphics[width=\textwidth]{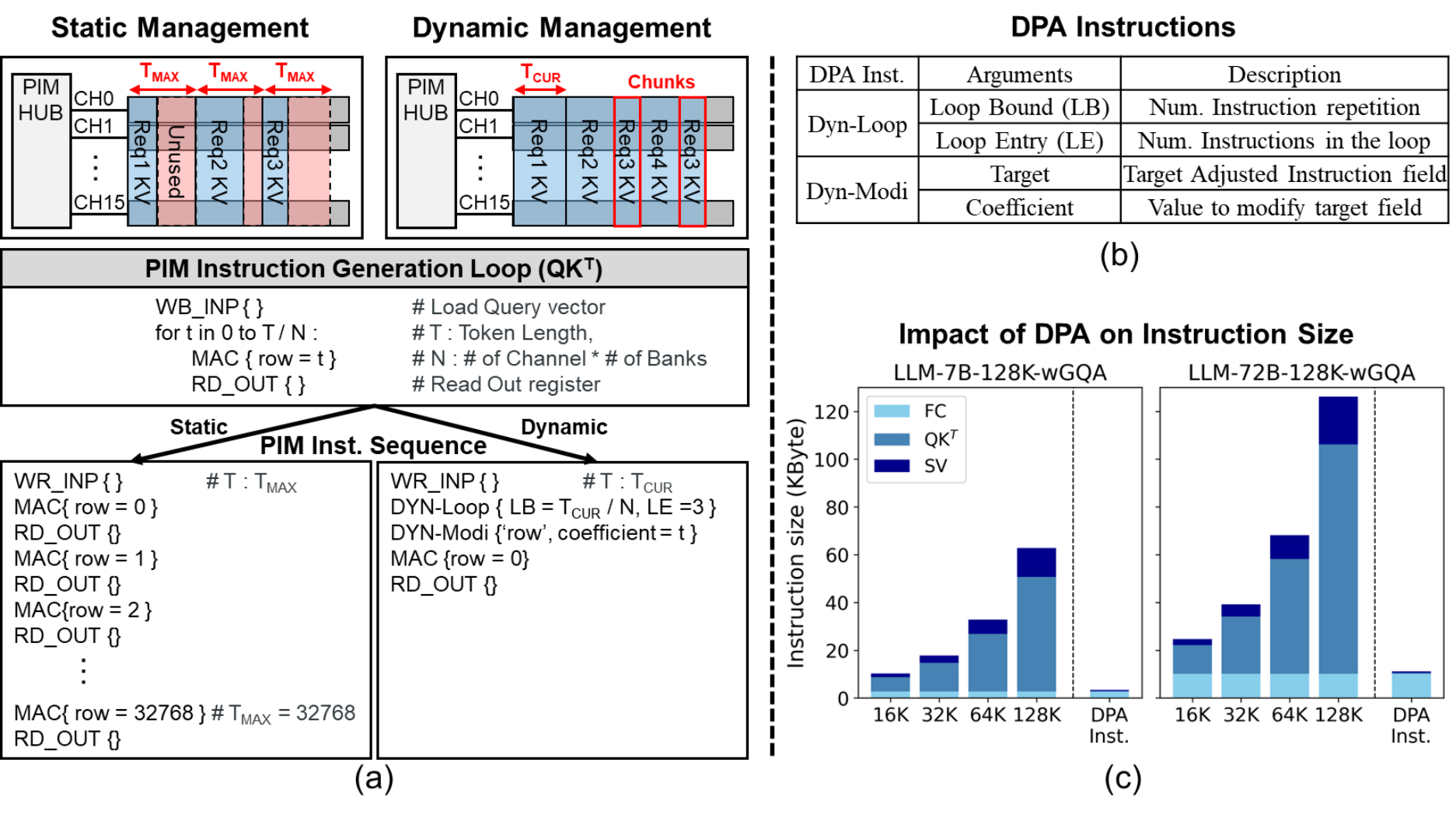}
        \vspace{-20pt}
        \caption{(a) \textcolor{black}{Limitations of static memory management in PIM where memory is allocated based on max token size,} (b) new DPA Instructions introduced with \arch, and   (c) impact of DPA on instruction size.}
        \label{fig:dpa-left}
    \end{minipage}
    \hfill
    \begin{minipage}[b]{0.36\textwidth}
        \centering
        \includegraphics[width=\textwidth]{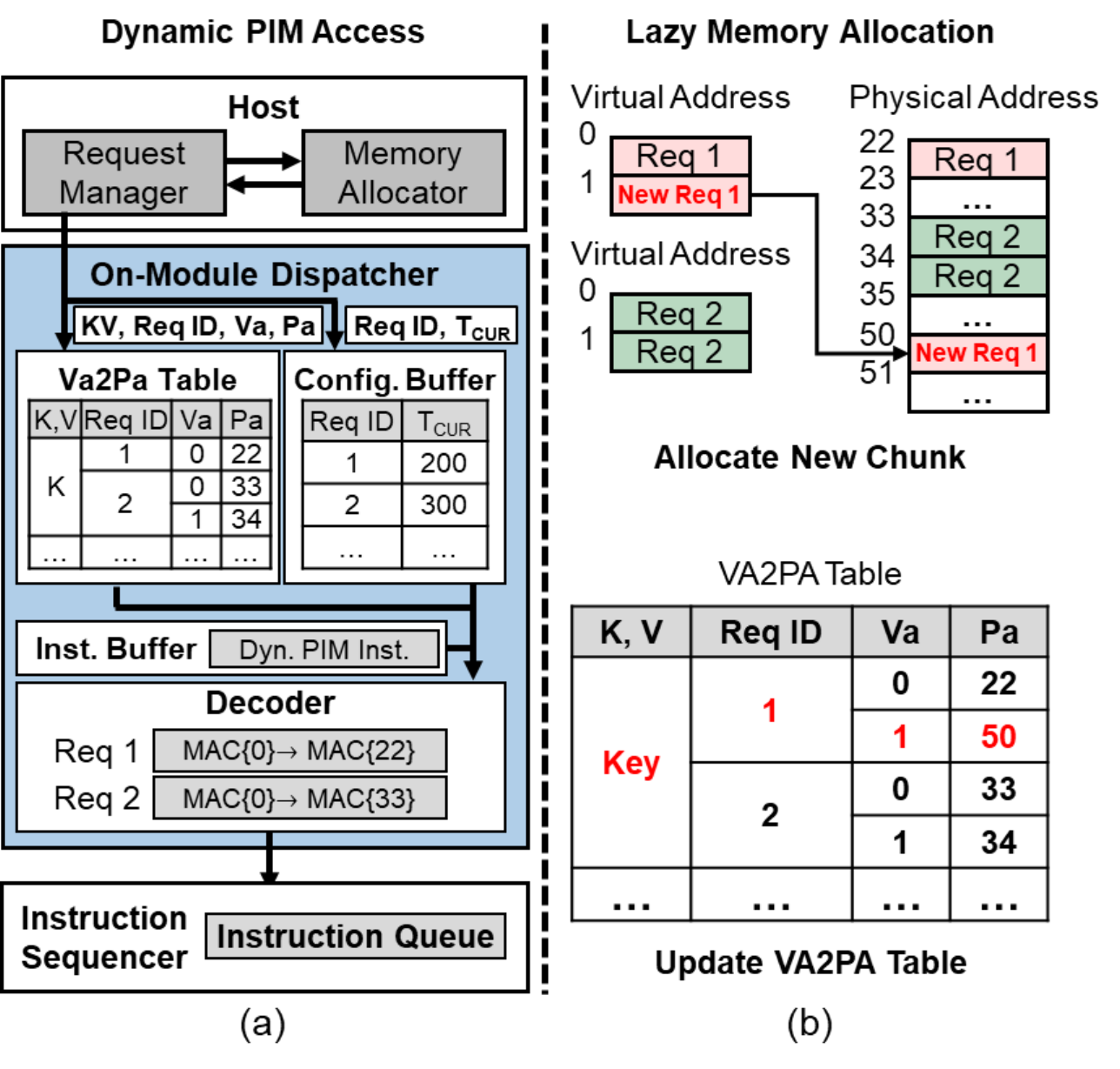}
        \vspace{-20pt}
        \caption{
        To enable dynamic memory management with DPA, high-level overview of 
        (a) on-module dispatcher and  (b) lazy memory allocation.}
        \label{fig:dpa-right}
    \end{minipage}
    \vspace{-10pt}
\end{figure*}

\textbf{DCS's Latency Saving.} 
DCS reduces latency by dynamically relaxing inter-command timing as dependencies permit, thereby increasing pipeline efficiency and enabling out-of-order issue of independent commands. Fig.~\ref{fig:pim-cmd-schedule}(d) provides an example where $W_0$ completes, the scheduler immediately issues $M_3$ because it is independent of $W_2$, even though $W_2$ has not yet finished (relaxing inter-command timing). This allows computation to proceed without waiting unnecessarily. Similarly, $M_7$ is issued before $R_6$ because they have no data dependency, which permits $M_7$ to directly follow $M_5$ with only a minimal $t_{CCDS}$ interval (out-of-order execution). By bypassing such stalls, the DCS scheduler improves MAC pipeline throughput and significantly reduces total execution time. In the example shown in Fig.~\ref{fig:pim-cmd-schedule}(d), DCS reduces the execution from 34 cycles to just 22 cycles.

\textbf{Enabling KV Cache Reuse in GQA.}
When operating across multiple DRAM rows, completing all computations on the currently opened row before moving to the next reduces ACT/PRE overhead; we refer to this as a \emph{row-reuse mapping}. When applied to GQA~\cite{ainslie2023gqa}, this means processing all inputs (queries/scores) that share the row-resident KV cache before switching DRAM rows. However, this mapping increases I/O transfer because queries/scores are frequently swapped into the GBuf, exacerbating pipeline stalls under \textit{static scheduling}. As shown in Fig.~\ref{fig:GQA_breakdown}, this raises DT-GBuf overhead even with weight reuse. DCS mitigates the issue by overlapping I/O transfers with MAC execution, reducing pipeline stalls and allowing \emph{row-reuse mapping} to yield performance gains. 

\section{Dynamic PIM Memory Management}
\label{sec:DPA}
Managing memory in long-context LLM decoding is challenging due to variability in context lengths. While GPU systems~\cite{pagedAttention} have implemented dynamic memory allocation to handle such variability, existing PIM architectures~\cite{cent, lee2021pimsw, kim2023samsung} remain fundamentally constrained by static control mechanisms. 
Conventional PIMs rely on a pre-compiled instruction sequence—a set of instructions used to execute a target operation—with fixed loop counts and operand addresses. This control scheme can handle repetitive computations, such as layer or head loops, whose iteration counts remain constant during inference. However, it cannot adapt to the variable-sized attention computation dictated by the current token length. Because all PIM instructions are statically compiled, the corresponding KV cache addresses must be pre-assigned at compile time, thereby preventing dynamic memory management during decoding. Thus, conventional PIMs do not provide the on-demand allocation required by vLLM’s paged-attention~\cite{pagedAttention}, leading to inefficient memory utilization.

To overcome this limitation, we introduce the \textit{Dynamic PIM Access} (DPA) controller, which effectively acts as a lightweight, pseudo-Memory Management Unit (MMU) for PIM. DPA supports diverse KV cache sizes and enables “lazy” memory allocation by providing a mechanism for runtime virtual-to-physical address translation—a capability absent in conventional PIM architectures.

\subsection{Conventional KV Cache Management in PIM}
\label{subsec:PIM-kv-cache-manage}
PIM-based systems accelerate Attention computations by managing the KV cache directly within their memory modules. However, existing systems rely on an inefficient static memory management scheme, as illustrated in Fig.~\ref{fig:dpa-left}(a). To handle the worst-case scenario, memory is preallocated based on the maximum possible token length. This leads to severe memory capacity underutilization when the actual token length is shorter than the reserved maximum, a common occurrence in dynamic long-context workloads.

The root cause of this underutilization is a fundamental architectural limitation: conventional PIM~\cite{AiMX-White-Paper} designs lack explicit support for control instructions that enable runtime-dependent operand or address computation. While some architectures (e.g.\!\cite{lee2021pimsw}) provide limited loop constructs via JUMP instructions, these loops can only iterate over statically pre-compiled physical memory addresses. As a result, even within a loop, memory accesses cannot be adjusted based on the actual token length observed at runtime. Consequently, the compiler is forced to conservatively generate instruction sequences that assume the worst-case maximum token length, which enforces rigid static memory allocation and prevents the system from adapting to dynamic context sizes during LLM decoding (Fig.~\ref{fig:dpa-left}(a)).

To overcome the limitations of static PIM memory, we leverage two key properties of the Attention computation structure. First, the instruction sequence exhibits a highly repetitive compute pattern (Fig.~\ref{fig:dpa-left}(a)), allowing it to be expressed as a compact loop instead of a long, static list. Second, and more critically, \textit{the instruction operands are token-length-dependent}. For example, the \texttt{MAC} row index is calculated directly from the current token length ($T_{cur}$) using the formula, $row=T_{cur}/(n_{CH}*n_{Bank})$. This enables the use of logical, loop-dependent virtual addresses (VA), which are translated to physical addresses (PA) at runtime using a VA2PA table.

\subsection{Dynamic PIM Access (DPA) Instructions}
\label{subsec:DynC}

Based on the insights of Attention's computational structure, we introduce Dynamic PIM Access (DPA) instructions, a flexible instruction set designed to avoid the static execution model of conventional PIMs. As outlined in Fig.~\ref{fig:dpa-left}(b), DPA instructions empower the PIM to handle dynamic workloads by encapsulating repetitive token operations into a compact, runtime-executable format. This is achieved through two key instructions:
\begin{itemize}
    \item \texttt{Dyn-Loop} instruction encodes a loop structure where the number of repetitions (Loop-Bound) is determined dynamically based on the request's actual token length, not a pre-defined maximum token length.
    \item \texttt{Dyn-Modi} instruction operates within this loop to adjust target operand fields (e.g., the row/col address for a \texttt{MAC} instruction) using a specified stride. This mechanism effectively generates a virtual address that is translated to a physical address on-the-fly, enabling the PIM to access dynamically allocated and non-contiguous KV cache memory—a capability impossible for conventional PIM controllers.
\end{itemize}
This dynamic and compact representation provides an advantage over prior PIM systems (~\cite{choi2024attacc,heo2024neupims,cent,AiMX-White-Paper}), which must generate instruction sequences whose size grows linearly with the token length. As shown in Fig.~\ref{fig:dpa-left}(c), this linear growth creates severe instruction buffer pressure and a scalability bottleneck for long-context inference. In comparison, DPA's encoding ensures the instruction size remains small and nearly constant regardless of context length, thereby avoiding command buffer bloat and enabling scalability to long context.

\subsection{On-Module PIM Instruction Dispatcher}
\label{subsec:dispatcher}
To support DPA, we enhance the PIM HUB (Fig.\ref{fig:System_Overview}(c)) with a lightweight on-module dispatcher that handles dynamic instruction generation and address translation. As shown in Fig.~\ref{fig:dpa-right}(a), this dispatcher consists of an instruction buffer to hold the compact, DPA-encoded instruction sequence; a configuration buffer for per-request information like request ID and token index ($T_{cur}$); and a Virtual-to-Physical Address (VA2PA) translation table to maintain the mapping between logical virtual addresses and the physical memory locations. The physical memory space is allocated in a granularity of \textit{chunk}, defined as \textit{channels} × \textit{banks} × \textit{rows}. A decoding unit combines the DPA-encoded instructions, the per-request information, and the resolved physical addresses from the VA2PA table to generate the final executable PIM instruction sequence. After decoding, the instructions are staged for execution in the Instruction Queue inside the PIM HUB's Instruction Sequencer (Fig.~\ref{fig:System_Overview}(a)).

Before LLM inference starts, the host initializes the dispatcher with per-request metadata, including the VA2PA table and the initial token length. Once assigned, the dispatcher manages token progression by incrementing the token length after each generation step, maintaining this state until the request completes. During this process, no host intervention is required. Host–PIM communication occurs only when a new request is assigned to a module, when an active request requires additional memory space, or when a completed request is released. 
As this communication does not occur on every decoding step, its overhead is negligible in practice. This localized update mechanism allows the dispatcher to efficiently handle variable-length decoding while minimizing communication latency between the host and PIM modules.

During instruction decoding, the Decoder reads operand addresses from the instruction buffer and resolves them through the VA2PA table according to the active request context. For example, as shown in Fig.~\ref{fig:dpa-right}(a), a MAC instruction with a virtual address of 0 can be dynamically translated to different physical addresses depending on the request ID—it resolves to address 22 for Request 1 but maps to address 33 for Request 2. This dynamic address translation enables the dispatcher to support non-contiguous and dynamically allocated memory regions across concurrent requests.
Importantly, instruction decoding is performed in a pipelined manner with instruction execution. Decoded instructions are staged into the instruction queue while previously decoded instructions are executed, allowing decoding and execution to proceed in parallel without introducing additional stalls on the critical path.

\textbf{Lazy Memory Allocation.} The integration of DPA and the on-module dispatcher enables efficient \textit{lazy memory allocation}, directly addressing memory underutilization in static PIM systems. Instead of reserving memory for worst-case scenarios, the host allocates 1MB chunks on demand as a request's KV cache grows, updating the VA2PA table accordingly. As shown in Fig.\ref{fig:dpa-right}(b), when more memory is needed, a new chunk is allocated and mapped non-contiguously via virtual addresses. This dynamic scheme reduces internal fragmentation to only the last-used chunk per request. Compared to static $T_{max}$ reservation, it significantly improves memory capacity efficiency (as discussed later in Sec.\ref{SubSec:Ablation}).

\section{\arch Implementation}
\label{sec:implementation}
\arch integrates TCP, DCS, and DPA through compiler and runtime techniques. To demonstrate its benefit, we apply our MLIR-based implementation to existing simulators for NeuPIMs~\cite{heo2024neupims} (NPU+PIM) and CENT~\cite{cent} (PIM-only), enabling evaluation across different types of PIM systems.

\begin{figure}[t]
    \centering
    \includegraphics[width=1\linewidth]{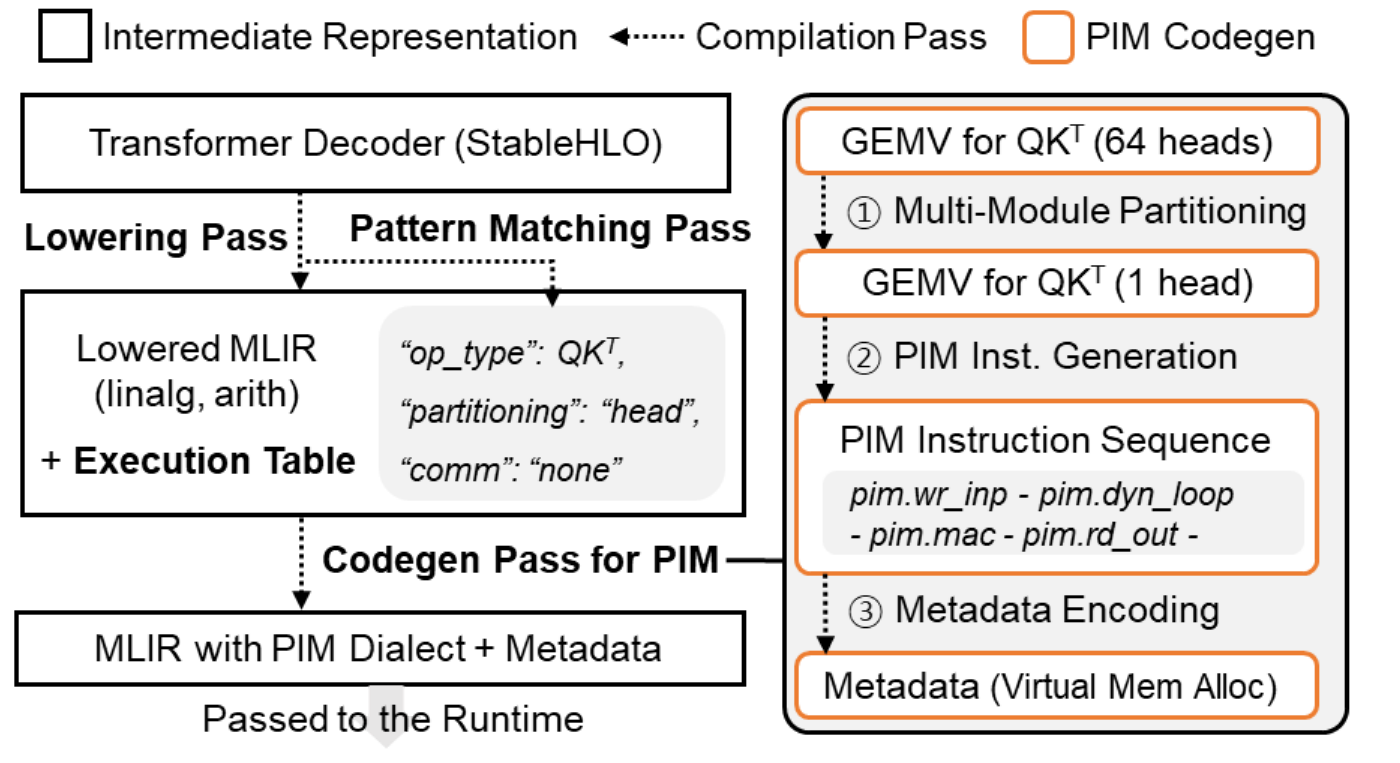}
    \vspace{-15pt}
    \caption{Overall compilation flow of \arch.}
    \label{fig:compilation}
    \vspace{-15pt}
\end{figure}

\subsection{Multi-PIM Compiler and Runtime}
\label{subsec:compiler}
We implement \arch on top of MLIR~\cite{mlir}, a flexible compiler framework designed for modular optimization across software and hardware layers (Fig.~\ref{fig:compilation}). By extending MLIR’s dialects, we generate PIM-specific code for Attention and feed-forward subgraphs in Transformer decoding workloads. \archp’s custom pattern-matching and code generation passes target PIM-amenable kernels (e.g., $QK^T$, $SV$, $FFN$), embedding dynamic partitioning and memory allocation metadata. Compilation is performed offline and does not affect inference latency.

For runtime support, we enhance the IREE~\cite{iree} runtime stack and its hardware abstraction layer (HAL) to interface with PIM SDKs provided by commercial platforms~\cite{AiMX-White-Paper}. This enables the deployment of \archp’s instruction sequences in realistic multi-node settings, where token-centric partitioning and dynamic memory management adaptively respond to variations of context length. These runtime extensions are implemented without assuming new hardware primitives, ensuring compatibility with NeuPIMs and CENT simulators.

\subsection{PIM System Integration}
To evaluate \arch in multi-node inference settings, we incorporate the compiler-generated PIM instruction sequences into simulators for both NeuPIMs and CENT. In NeuPIMs, \arch offloads bandwidth-bound Attention operations to PIM modules, while in CENT, all computations are handled with PIM. We apply \archp's hardware-aware optimizations—including on-module dispatch logic and I/O buffering—within the simulation backends of both platforms, with DRAM command timing and resource contention calibrated with AiMX~\cite{AiMX-White-Paper} PIM specifications.

\subsection{Hardware Overhead}
\label{subsubsection::Hardware Overhead}
\archp’s proposed hardware modifications are lightweight, as the I/O-aware buffering incurs a minimal area overhead, occupying just 0.47\% of the MAC unit area~\cite{AiM_ISSCC} per PIM bank, estimated by synthesizing the output buffer logic using CACTI~\cite{CACTI}. 
In addition, enabling DCS within PIM controllers in PIM HUB~\cite{AiMX-Hotchips-2022,AiMX-Hotchips-2023,AiMX-Hotchips-2024} adds only a negligible controller cost: across all HUB control blocks~\cite{AiMX-White-Paper}, the area increases by 0.5\% and the power by 1.3\%. This overhead accounts for the additional dependency-tracking structures and logic required by DCS, including a per-controller 576B metadata table (D-Table and S-Table) and the associated dependency-check unit used to validate per-entry hazards.
The on-module dispatcher supporting DPA incurs a 4\% area overhead. This dispatcher requires less than 200KB for all its internal buffers (VA2PA, command, configuration), which is much smaller than the 512KB GPR capacity in a typical PIM HUB, ensuring efficient integration without significant area pressure.

\begingroup
\renewcommand{\arraystretch}{1.2} 
\begin{table}[t]
\centering
\caption{\arch module configurations.}
\resizebox{\linewidth}{!}
    {
    \fontsize{20pt}{22pt}\selectfont
    \begin{tabular}{c|c|c}
    \toprule
    \cline{1-3}
    \multirow{3}{*}{NeuPIMs~\cite{heo2024neupims}} & Compute & 8 Matrix Units(256 TFLOPS), 32 PIM channels\\
    \cline{2-3}
                       & Memory & 32GB \\
                        \cline{2-3}
                       & Internal BW & 32TB/s \\
                        \cline{2-3}
    \cline{1-3}
    \multirow{3}{*}{CENT~\cite{cent}}                        & Compute & PNM (3 TFLOPS), 32 PIM channels \\
    \cline{2-3}
                                            & Memory & 16GB  \\
                                            \cline{2-3}
                                            & Internal BW & 16TB/s \\
    \cline{1-3}
    \bottomrule
    \end{tabular}
    }
    \label{table:system_config}
    \vspace{-10pt}
\end{table}
\endgroup

\begin{figure*}[!t]
    \centering
    \includegraphics[width=\textwidth]{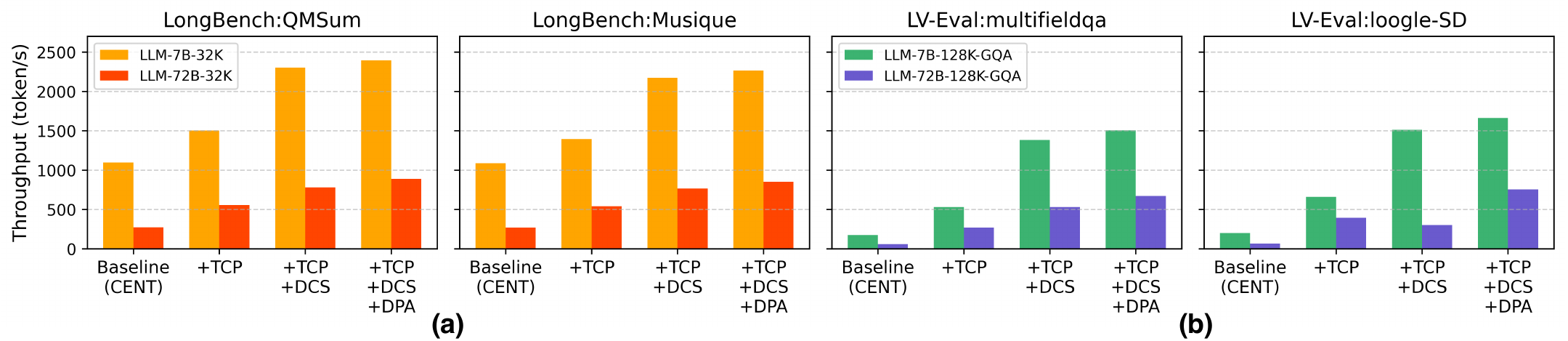}
    \vspace{-15pt}
    \caption{Throughput of PIM-only systems — 7B: 8 modules (128GB), 70B: 32 modules (512GB). (a) Non-GQA LLMs on LongBench. (b) GQA-enabled LLMs on LV-Eval. Each bar shows improvements from TCP, DCS, and DPA, using optimal TP/PP settings.}
    \label{fig:main_cent}
    \vspace{-10pt}
\end{figure*}

\begin{figure*}[!t]
    \centering
    \includegraphics[width=\textwidth]{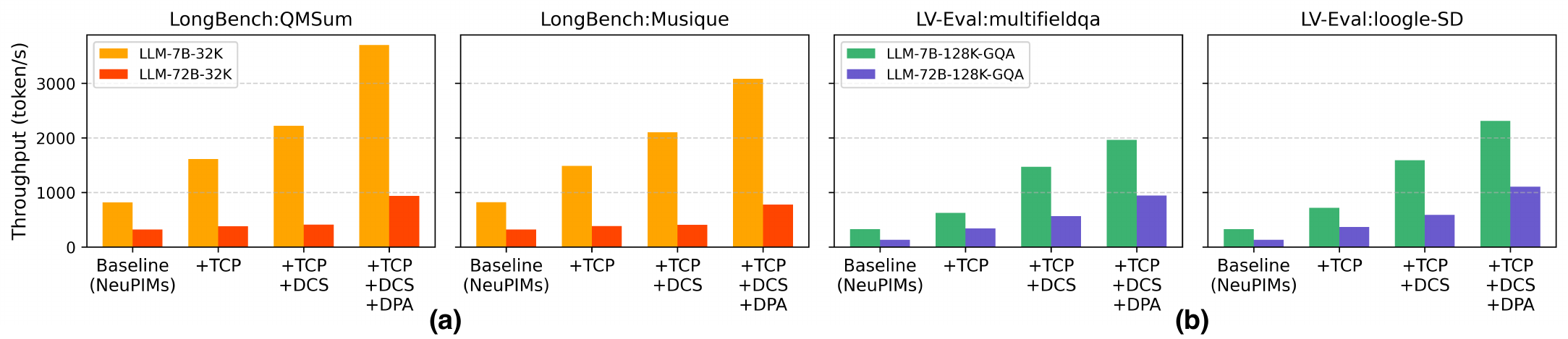}
    \vspace{-15pt}
    \caption{Throughput of xPU-PIM systems — 7B: 4 modules (128GB), 70B: 16 modules (512GB). (a) Non-GQA LLMs on LongBench. (b) GQA-enabled LLMs on LV-Eval. Each bar shows improvements from TCP, DCS, and DPA, using optimal TP/PP settings.}
    \label{fig:main_npupim}
    \vspace{-10pt}
\end{figure*}

\section{Evaluation}

\subsection{Evaluation Settings}
\label{Sec:Evaluation}
We evaluate \arch using a range of LLMs (Table~\ref{tab:LLMs}) with context lengths up to 128K, across four tasks from LongBench~\cite{bai2024longbench} and LV-Eval~\cite{lveval} (Table~\ref{tab:longbench_statics}). Non-GQA models (LLM-7B/72B-32K) are tested with LongBench, while GQA-enabled models (LLM-7B/72B-128K) are evaluated with LV-Eval. For comparison, we use two PIM-based baselines: CENT~\cite{cent}, a PIM-only system with 16GB per module, and NeuPIMs~\cite{heo2024neupims}, a hybrid xPU+PIM system with 32GB per module. We modify both simulators~\cite{cent,heo2024neupims} to integrate our techniques, configuring 128GB for 7B and 512GB for 72B models, following prior PIM studies. \arch is modeled using a validated Ramulator-based simulator incorporating AiMX~\cite{AiMX-Hotchips-2022,AiMX-Hotchips-2023,AiMX-Hotchips-2024} architecture, and is evaluated using the parameters detailed in Table~\ref{table:system_config}.

\begin{figure}[t]
    \centering
    \includegraphics[width=1\columnwidth]{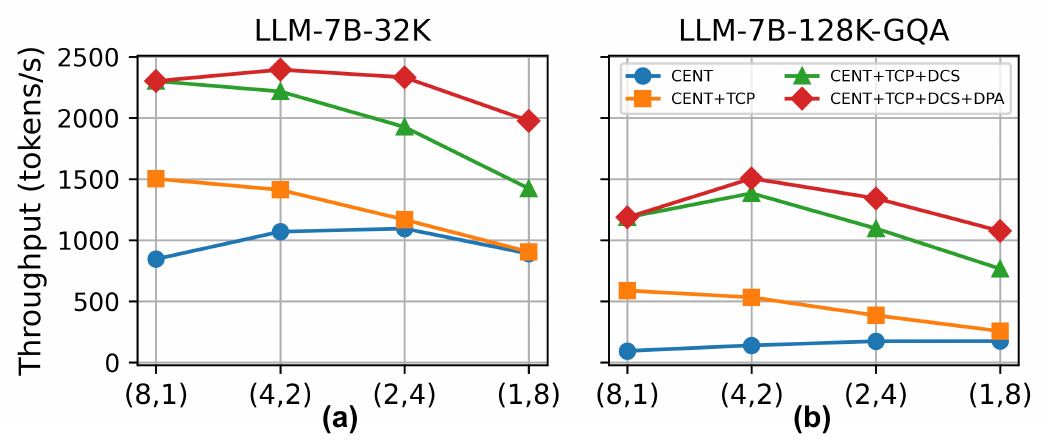}
    \vspace{-20pt}
    \caption{Throughput of various (TP,PP) on (a) LLM-7B-32K with LongBench QMSum (b) LLM-7B-128K-GQA with multifieldqa.}
    \label{fig:tradeoff}
    \vspace{-10pt}
\end{figure}

\subsection{Performance Evaluation.}
\label{subsec:perf-cmpr}
We evaluate the end-to-end throughput improvements of our proposed techniques across both PIM-only (CENT~\cite{cent}) and heterogeneous xPU+PIM systems (NeuPIMs~\cite{heo2024neupims}). As shown in Fig.~\ref{fig:main_cent} and \ref{fig:main_npupim}, we incrementally apply three optimizations of \arch: token-centric PIM partitioning (TCP), dynamic PIM command scheduling (DCS), and dynamic PIM access (DPA).

\begin{figure}[t]
    \centering
    \includegraphics[width=1.0\columnwidth]{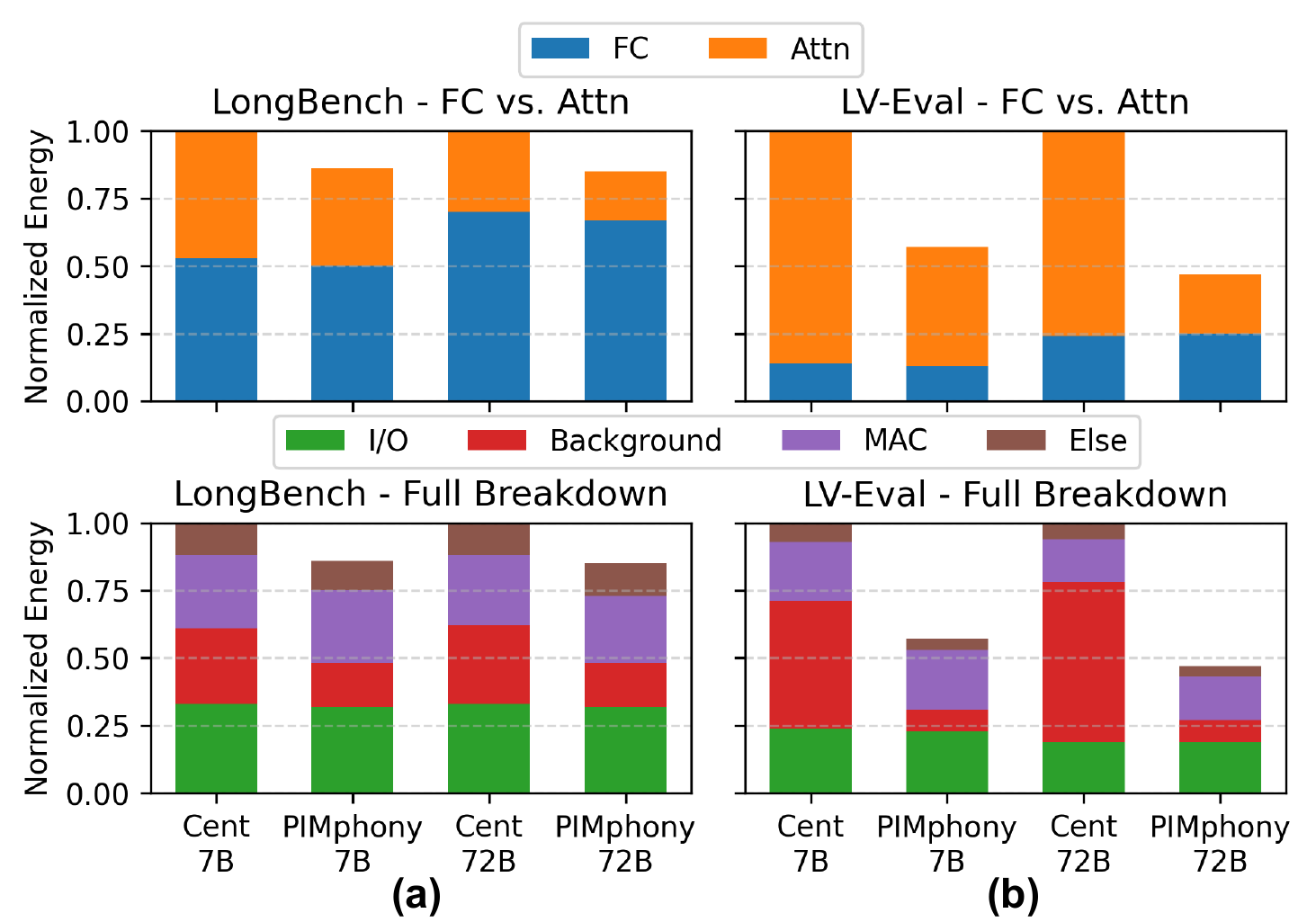}
    \vspace{-15pt}
    \caption{
    Energy breakdowns of CENT vs. CENT+PIMphony for 7B and 72B models: (a) non-GQA models evaluated on LongBench (32K), (b) GQA models evaluated on LV-Eval (128K). Top: FC and Attention. Bottom: MAC, I/O, Background, and Else.
    }
    \label{fig:energy_breakdown}
    \vspace{-10pt}
\end{figure}

\textbf{Overall Throughput Gains.}
Performance consistently improves with the progressive application of the proposed methods. In Fig.~\ref{fig:main_cent}(a) and~\ref{fig:main_npupim}(a), we observe a 2.1-4.5x speedup for GQA-disabled models (LLM-7B-32K, LLM-72B-32K) evaluated on LongBench. For GQA-enabled models (LLM-7B-128K, LLM-72B-128K) evaluated on LV-Eval (up to 128K tokens), our methods achieve up to 11.3× speedup, demonstrating their strong effectiveness in large-context scenarios. Notably, PIMphony’s relative throughput gain is often higher for the 72B models because larger models suffer more from baseline inefficiencies (e.g., PIM underutilization in the CENT system), creating greater room for improvement. While this trend is evident in PIM-only systems, the advantage in xPU+PIM systems becomes more pronounced under very long-context workloads (e.g., 128K), where PIM-side execution increasingly dominates. 

\textbf{Effectiveness of TCP, DCS, and DPA.}
TCP directly addresses the poor channel utilization in PIM-only and xPU+PIM systems, which is caused by the small batch sizes and context imbalances inherent to long-context workloads, more pronounced in longer context (LV-Eval). While TCP effectively boosts PIM utilization, the system can become I/O-bound, a problem that DCS resolves by overlapping computation and data movement. This synergy is especially impactful for GQA-enabled models, where DCS unlocks the gains of \emph{row-reuse mapping} by overlapping the additional I/O transfers (especially \texttt{WR-INP}) with \texttt{MAC} execution, thereby amplifying ACT/PRE savings over TCP-only configurations.
Finally, in heterogeneous xPU+PIM systems, DPA plays a critical role in maintaining system-level balance. As TCP and DCS accelerate the PIM-bound Attention stage, the xPU-bound FC stage increasingly becomes the dominant performance bottleneck. DPA alleviates this imbalance by enabling larger batch sizes, which in turn improves NPU utilization and maximizes end-to-end throughput.

\subsection{Ablation Study}
\label{SubSec:Ablation}
In this section, we evaluate the effectiveness of PIMphony across several key metrics: multi-PIM parallelization, latency and energy breakdowns, system and context length scalability, memory capacity utilization, and a direct throughput comparison with GPU systems. Unless otherwise noted, we present results primarily from the PIM-only system to more distinctively illustrate the impact of PIMphony on PIM itself.

\begin{figure}[t]
    \centering
    \includegraphics[width=1.0\columnwidth]{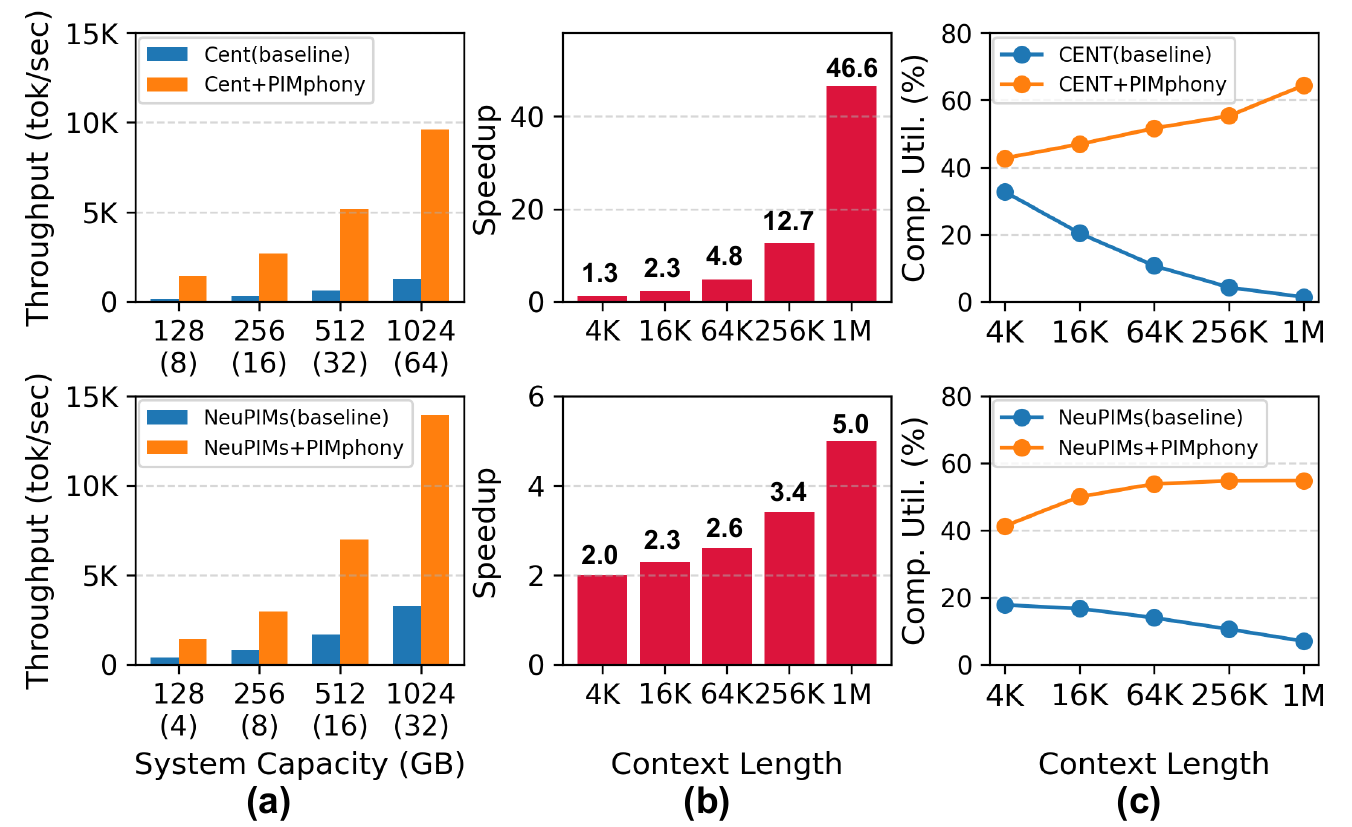}
    \vspace{-15pt}
    \caption{Scalability of PIMphony on LLM-7B-128K-wGQA with 3-sigma context variation for CENT (Top) and NeuPIMs (Bottom). (a) At 64K context, throughput scales with capacity (128–1024GB; 8–64 modules for CENT, 4–32 for NeuPIMs). (b)(c) At 512GB, PIMphony outperforms baselines as context length increases from 4K to 1M.}
    \label{fig:scalability}
    \vspace{-10pt}
\end{figure}

\textbf{Tensor vs. Pipeline Parallelization. }
Fig.~\ref{fig:tradeoff} illustrates the impact of \arch's techniques on multi-node PIM parallelization, evaluated on CENT. We incrementally apply TCP, DCS, and DPA to assess their cumulative benefits. For LLM-7B-32K (woGQA), TCP significantly enhances TP efficiency by mitigating channel underutilization. DCS further improves Attention performance by reducing I/O idle time, while DPA increases the effective batch size, enabling PP to deliver marginally higher throughput. For LLM-7B-128K (wGQA), DCS shows a more pronounced effect due to the intensified input transfer pressure from GQA. Additionally, KV cache reuse from GQA, combined with DPA, boosts batch size and further motivates PP, achieving 20\% higher throughput improvement. These results demonstrate the effectiveness of \arch in maximizing multi-PIM system efficiency.


\textbf{Energy Breakdown.}
PIMphony’s energy efficiency stems from its dramatic reduction of runtime-dependent background energy. As shown in Fig.~\ref{fig:energy_breakdown}, the baseline's low MAC utilization causes background energy to constitute a staggering 71.5\% of the total energy in Attention layers. By accelerating Attention execution by up to 19$\times$, PIMphony drastically cuts the runtime, which in turn slashes the background energy's share to just 13.0\% and achieves up to a 3.46$\times$ reduction in Attention. These gains are amplified in GQA-enabled models, as their higher speedup potential leads to an even greater reduction in runtime and, consequently, background energy.


\textbf{Scalability with System Capacity and Context Length. }
PIMphony scales robustly with both system capacity and context length, widening its lead over baselines. Throughput improves with capacity (128GB–1024GB), confirming efficient module use (Fig.\ref{fig:scalability}(a)). Benefits are greater when scaling context length to 1M tokens on a fixed 512GB system. CENT collapses under pipeline bubbles, dropping to 2\% utilization, while PIMphony achieves 46.6$\times$ speedup. NeuPIMs scale more stably via tensor parallelism, yet PIMphony still delivers 5.0$\times$ speedup (Fig.\ref{fig:scalability}(b)). Fig.~\ref{fig:scalability}(c) explains this trend: PIMphony makes Attention more efficient than FC layers, so longer contexts—where Attention dominates—boost system utilization, unlike baselines where bottlenecks worsen. Importantly, these gains are not confined to long contexts; even at short contexts (e.g., 256 tokens), PIMphony achieved a 2.1$\times$ speedup over the baselines.


\begin{figure}[t]
    \centering
    \includegraphics[width=1.0\columnwidth]{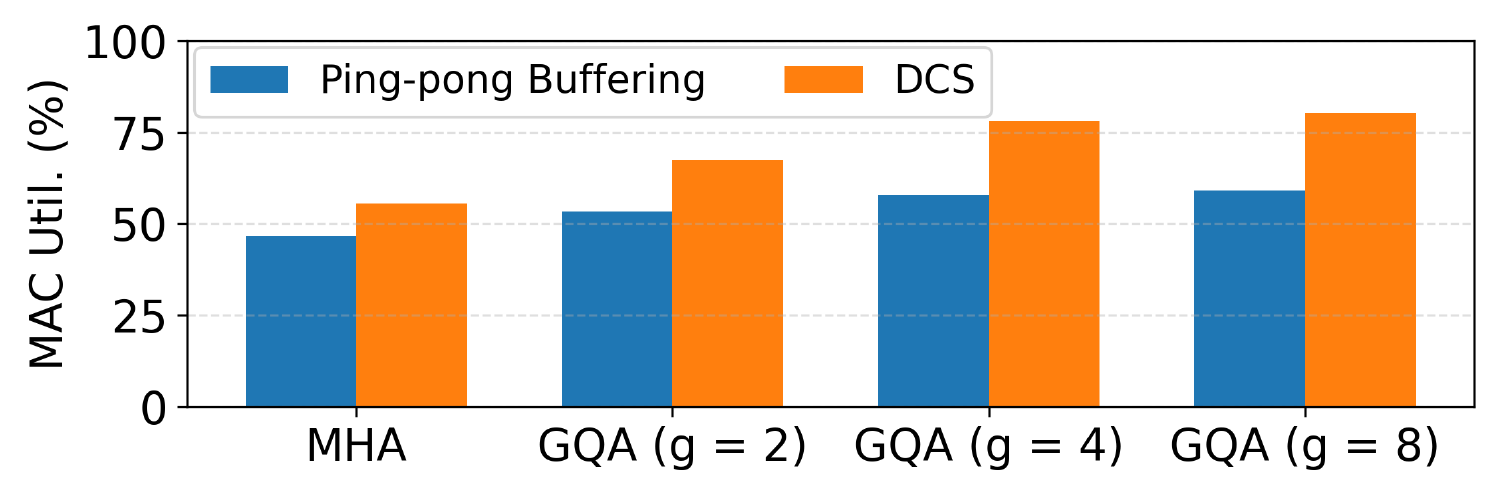}
    \vspace{-15pt}
    \caption{Compute utilization for Attention operations comparing \emph{ping-pong buffering} and DCS. X-axis: MHA and GQA with group size $g\in{2,4,8}$. Both apply the \emph{row-reuse mapping} in GQA.}
    \label{fig:DCS_Pingpong}
    \vspace{-10pt}
\end{figure}

\begin{figure}[t]
    \centering
    \includegraphics[width=1.0\columnwidth]{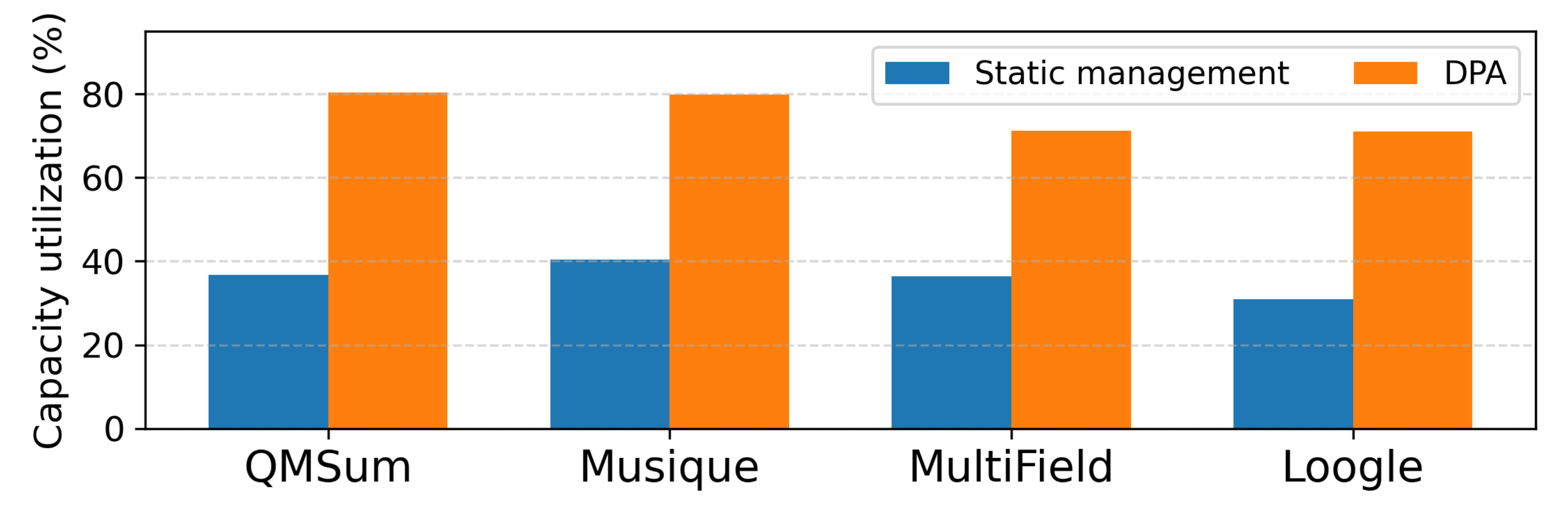}
    \vspace{-15pt}
    \caption{Capacity utilization with and without DPA in PIMphony. QMSum/Musique use 7B-32K; MultiField/Loogle use 7B-128K.
    }
    \label{fig:capacity_util}
    \vspace{-10pt}
\end{figure}

\textbf{DCS vs. Ping-pong Buffering} Prior works\cite{PipePIM,DH-PIM} adopt \emph{ping-pong buffering}: a single buffer is split into two regions so two operations that touch disjoint regions can overlap—thus, like DCS, it can overlap I/O transfers (\texttt{WR-INP}, \texttt{RD-OUT}) with \texttt{MAC} to improve MAC utilization. However, because \emph{static scheduling} does not know true data dependencies at the entry level, I/O transfers and \texttt{MAC} cannot access the same region concurrently to avoid data hazards; effectively, overlap is restricted to different regions. Consequently, when I/O transfers and \texttt{MAC} must switch which region they operate on, the swap can occur only after both regions become idle, which introduces hand-off pipeline stalls. In contrast, DCS keeps one buffer, tracks per-entry dependencies, and relaxes inter-command timing to enable entry-level overlap within the same buffer without hand-offs, yielding a longer, stall-resistant pipeline. With the same total buffer size, our ping-pong baseline for I/O hiding exhibits hand-off pipeline stalls—exacerbated by shorter pipelines from smaller per-region buffers—whereas DCS sustains overlap; across Attention settings, DCS achieves up to 1.4$\times$ higher compute-unit utilization (Fig.~\ref{fig:DCS_Pingpong}).

\begin{figure}[t]
    \centering
    \includegraphics[width=1.0\columnwidth]{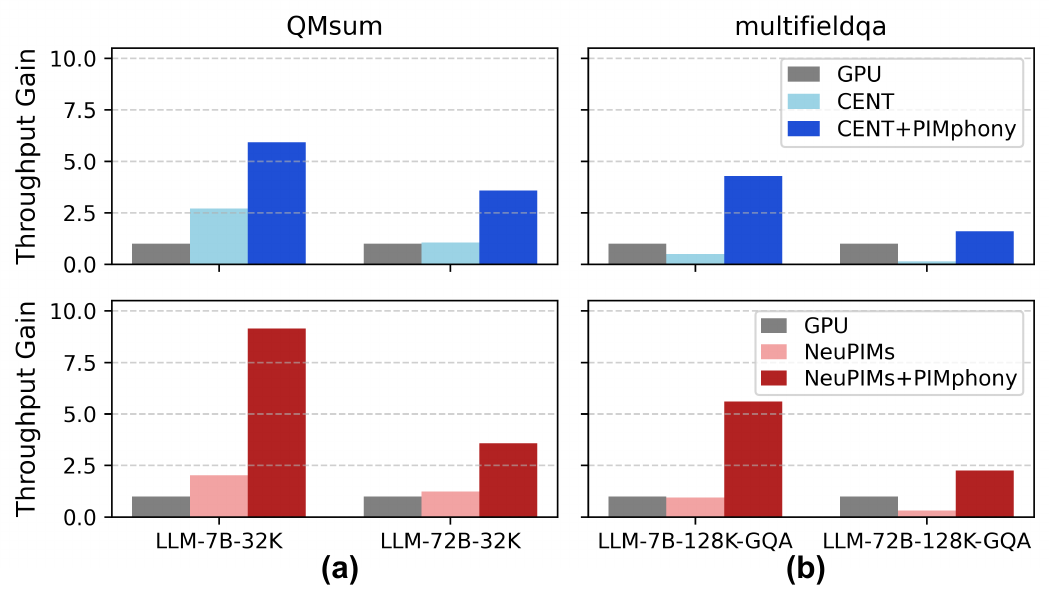}
    \vspace{-15pt}
    \caption{Throughput comparison of GPU and \arch. (a) Non-GQA LLM on QMSum (LongBench). (b) GQA-enabled LLM on multifieldqa (LV-Eval). For fairness, GPU memory is matched to PIMphony: two A100-80GB for LLM-7B, eight for LLM-72B.}
    \label{fig:gpu-throughput}
    \vspace{-5pt}
\end{figure}

\textbf{Impact of DPA on Capacity Utilization.}
To assess DPA’s capacity efficiency, we conduct an ablation comparing static memory management with DPA’s dynamic approach. As shown in Fig.~\ref{fig:capacity_util}, static memory management severely underutilizes memory—capacity utilization ranges from 31.0\% to 40.5\% across workloads—due to coarse-grained reservations sized for maximum token lengths, which over-provision capacity. DPA instead uses lazy, on-demand allocation of non-contiguous chunks at runtime. Although this introduces minor fragmentation limited to each request’s final chunk, it removes maximum-length reservations and markedly improves efficiency. Empirically, DPA raises average capacity utilization to 75.6\%, more than doubling the static baseline.

\textbf{Throughput Comparison with GPU System.}
We compare PIMphony-enabled systems with a strong GPU baseline—A100s with flash-decoding (FD)\cite{flashdecoding2023} and paged-attention (PA)\cite{pagedAttention}—using memory-matched configurations (Fig.~\ref{fig:gpu-throughput}). PIMphony delivers substantial speedups, especially on non-GQA models with high memory demand. On larger 72B models, the GPU’s advantage on compute-intensive FC layers narrows PIMphony’s relative gains. Even so, PIMphony shows a key strength on GQA workloads: while KV cache reuse benefits GPUs, it creates an I/O bottleneck for PIMs due to repeated input transfers. Our DCS technique mitigates this by overlapping transfers with computation, hiding the overhead and unlocking significant gains even in workloads traditionally favorable to GPUs.

\section{Related Work}
\label{Sec:RelatedWork}

\textbf{Multi-Node LLM Serving.} 
As LLMs scale, their memory and bandwidth demands grow, making parallelization—such as tensor~\cite{shoeybi2019megatron} and pipeline parallelism~\cite{huang2019gpipe}—crucial for serving. Modern systems~\cite{TensorRT-LLM,SGLang, vLLM} further improve GPU utilization with continuous batching~\cite{yu2022orca} and efficient KV cache management~\cite{pagedAttention}. Still, GPU underutilization persists during token generation~\cite{patel2024splitwise, he2024fastdecode}.
This paper targets improved PIM utilization for long-context LLM inference, where GPUs face bandwidth and capacity bottlenecks. We propose architectural and runtime techniques—token-centric partitioning, dynamic command scheduling, and dynamic memory management—that boost MAC and memory efficiency in both PIM-only and xPU+PIM systems.

\textbf{DRAM-Based PIM.}
DRAM-based Processing-in-Memory (PIM) is a key solution for accelerating LLMs, with general-purpose platforms like UPMEM-PIM~\cite{Mutlu.2022,Mutlu.20227oi,Rhu.2023} and domain-specific accelerators such as HBM-PIM~\cite{lee2021hardware,kwon202125,kim2022aquabolt,kim2023samsung, kim2024breakthrough} and AiM~\cite{AiM_ISSCC,AiM_JSSC,he2020newton}. These are optimized for tasks like matrix multiplication. Recent work addresses limitations in both homogeneous~\cite{Rosing.2022,kal2023aespa} and heterogeneous xPU–PIM systems~\cite{choi2024attacc,heo2024neupims, seo2024ianus, li2024specpim, wang2022cnn}. AiMX~\cite{AiMX-Hotchips-2022,AiMX-Hotchips-2023,AiMX-Hotchips-2024,AiMX-White-Paper}, which integrates FPGA with AiM, further advances compute and parallelism for LLMs, signaling a shift toward specialized, efficient hardware.

\textbf{DRAM-Based PIM for LLM Inference.}
DRAM-based PIM architectures are gaining attention for accelerating memory-bound phases of LLM inference by leveraging high internal bandwidth near memory arrays. Recent work explores compiler co-optimization, dynamic scheduling, and CXL-based system integration to support complex model execution on PIM~\cite{kim2024pimdl,papi2025,pimnet2025,toleo2024}, highlighting a shift toward holistic hardware–software co-design for LLM acceleration.

\textbf{PIM-Based Memory Management Units.}
Only limited prior work~\cite{PIM-MMU} has examined memory management support in Processing-in-Memory architectures. The most relevant effort is the MMU design proposed for the UPMEM system~\cite{Mutlu.20227oi}, which provides an address translation facility primarily to manage data transfers between DRAM and each PIM core’s local SRAM. This mechanism operates similarly to a DMA engine and is optimized for orchestrating bulk data movement across the DRAM–SRAM boundary. However, its scope is restricted to fixed data-transfer pipelines and does not address the challenges associated with dynamic memory allocation or operand remapping within DRAM itself.
In contrast to these prior approaches, our work performs virtual-to-physical address translation directly inside the PIM. This design enables on-demand address resolution and dynamic KV cache allocation within DRAM-based PIM modules, thereby supporting the flexible memory management required by modern LLM inference systems.

\textbf{Digital Accelerators for LLM Inference}
\label{subsub:accelerator}
Recent digital accelerator studies~\cite{hybe, specboost,mecla,hong2023dfx,accllm} have explored various architectural optimizations to enhance transformer-based LLM inference efficiency. Among them, Hybe~\cite{hybe} directly addresses the long-context LLM inference caused by the imbalance between compute-intensive prefill and memory-intensive decode stages. It adopts a GPU–NPU hybrid architecture, using GPUs for the prefill stage and lightweight NPUs for the decode stage to maximize hardware utilization. In addition, Hybe employs fine-grained KV transmission and stage-wise pipelining to reduce GPU memory pressure and minimize hardware stalls.

\section{Conclusions}
We propose \arch, a multi-node PIM acceleration system featuring an MLIR-based compiler and novel techniques for partitioning, scheduling, and memory management. By systematically resolving key bottlenecks, \arch significantly enhances throughput for long-context LLM inference up to 11.3$\times$ and 8.4$\times$ on PIM-only and xPU+PIM systems, respectively.


{\section*{Acknowledgment}
This work was partly supported by the National Research Foundation of Korea (NRF) grant funded by the Korea government (MSIT)
(No. RS-2023-00260527 and No. RS-2025-00561961). This work was also partly supported by the Institute of Information \& Communications Technology Planning \& Evaluation (IITP) grant funded by the Korea government (MSIT) (No. 2022-0-00971, Logic Synthesis for NVM-based PIM Computing Architecture, and No. RS-2020-II201373, Artificial Intelligence Graduate School Program (Hanyang University)). This work was also partly supported by SK Hynix Inc.

\bibliographystyle{IEEEtranS}
\bibliography{refs}

\end{document}